\documentclass[
trackchanges,
twocolumn
]{aastex701}

\usepackage{amsmath}
\usepackage{chngcntr}

\begin{document}

\title{Revisiting Disk Winds in Active Galactic Nuclei as an Origin of Cosmic Gamma-ray and Neutrino Backgrounds}

\correspondingauthor{Nobuyuki Sakai}

\author[orcid=0009-0004-5978-1785,sname='Sakai']{Nobuyuki Sakai}
\affiliation{Department of Earth and Space Science, Graduate School of Science, The University of Osaka, 1-1 Machikaneyama, Toyonaka, Osaka 560-0043, Japan}
\email[show]{u938638f@ecs.osaka-u.ac.jp} 

\author[orcid=0000-0002-7272-1136,gname=Yoshiyuki, sname='Inoue']{Yoshiyuki Inoue}
\affiliation{College of Systems Engineering and Science, Shibaura Institute of Technology, 307 Fukasaku, Minuma-ku, Saitama City, Saitama 337-8570, Japan}
\affiliation{Department of Earth and Space Science, Graduate School of Science, The University of Osaka, 1-1 Machikaneyama, Toyonaka, Osaka 560-0043, Japan}
\affiliation{Interdisciplinary Theoretical \& Mathematical Science Center (iTHEMS), RIKEN, 2-1 Hirosawa, 351-0198, Japan}
\affiliation{Kavli Institute for the Physics and Mathematics of the Universe (WPI), UTIAS, The University of Tokyo, 5-1-5 Kashiwanoha, Kashiwa, Chiba 277-8583, Japan}
\email{yoshiyuki.inoue.sci@osaka-u.ac.jp}

\author[orcid=0000-0003-1052-6439,gname=Ellis, sname='Owen']{Ellis R. Owen}
\affiliation{Astrophysical Big Bang Laboratory (ABBL), RIKEN Pioneering Research Institute (PRI), Wako, Saitama 
351-0198, Japan}
\affiliation{Department of Earth and Space Science, Graduate School of Science, The University of Osaka, 1-1 Machikaneyama, Toyonaka, Osaka 560-0043, Japan}
\email{ellis.owen@riken.jp}

\begin{abstract}

The origin of the cosmic neutrino background (CNB) and the cosmic gamma-ray background (CGB) remains uncertain. 
Accretion disk winds driven by active galactic nuclei (AGNs) have been proposed as possible contributors, but their predicted background levels depend sensitively on poorly constrained wind energetics and ambient densities.
We revisit the AGN disk-wind scenario by constructing a lepto-hadronic wind model calibrated with radio and GeV gamma-ray data of nearby \textit{Fermi}-LAT-detected Seyfert galaxies. 
In our framework, cosmic rays accelerated both at wind-driven forward and reverse shocks produce synchrotron, external-Compton, and hadronic emission. 
We also incorporate recent XRISM constraints on wind parameters. 
Applying our calibrated lepto-hadronic models to an AGN population synthesis model, we find that disk winds contribute at most $\lesssim 5\%$ of the CGB above 10~GeV and $\lesssim 10\%$ of the CNB around 100~TeV, suggesting that they are unlikely to dominate both backgrounds. 
Finally, we identify nearby Seyfert galaxies hosting ultrafast outflows as promising targets for future TeV gamma-ray and TeV--PeV neutrino observations, which would offer firm tests of the disk-wind scenario.

\end{abstract}



\section{Introduction}\label{sec:intro}

The cosmic neutrino background (CNB) is an isotropic flux of high-energy neutrinos from extragalactic sources.  
The IceCube Neutrino Observatory has detected the CNB \citep[e.g.,][]{2015ApJ...809...98A} and recently reported improved measurements in the $1~\mathrm{TeV}$--$100~\mathrm{PeV}$ range \citep{2025arXiv250722233A, 2025arXiv250722234A}.  
KM3NeT \citep{2016JPhG...43h4001A} also reported a cosmic neutrino candidate with an energy of $\sim100$~PeV \citep[][although this has not been confirmed by IceCube; see  \citealt{IceCube2025PhRvL.135c1001A}]{2025Natur.638..376K}. 
These high-energy neutrinos are thought to mainly originate in astrophysical environments via hadronuclear (\textit{pp}) interactions and/or photo-pion ($p\gamma$) interactions.   

A variety of sources have been proposed as the origin of the CNB, such as blazars \citep[e.g.,][]{1992A&A...260L...1M,2001PhRvL..87v1102A, 2014PhRvD..90b3007M,2015MNRAS.452.1877P, Rodrigues2018ApJ...854...54R}, star-forming galaxies \citep[e.g.,][]{2013PhRvD..87f3011H, 2015ApJ...806...24S, 2018PASJ...70...49S, 2024MNRAS.529.4137C, 2025arXiv250418721O}, and coronae of active galactic nuclei \citep[AGNs, e.g.,][]{1991PhRvL..66.2697S, 2013PhRvD..88d7301S, 2015JETP..120..541K, 2019ApJ...880...40I, Murase2020PhRvL.125a1101M, Fiorillo2025ApJ...989..215F}.  
However, stacking analyses of blazars imply their contribution to the TeV--PeV CNB flux is limited to $\lesssim30\%$ \citep[e.g.,][]{2017ApJ...835...45A, 2020ApJ...890...25Y}. 
Furthermore, using jet models, \citet{2015MNRAS.451.3649J} showed that X-ray-inferred AGN populations yield neutrino fluxes well above limits set by IceCube.
Studies have also demonstrated that if star-forming galaxies explain the entire CNB flux, their gamma-ray emission would overshoot the observed cosmic gamma-ray background \citep[CGB; e.g.][]{2013PhRvD..88l1301M, 2017ApJ...836...47B}. 
Moreover, since neutrino production in coronae has been shown to be highly dependent on source properties \citep{Inoue2024PASJ...76..996I, Lemoine2025A&A...697A.124L, 2025arXiv250706110S}, their contribution to the CNB is uncertain.

AGN disk winds have been discussed as another possible source of background neutrinos \citep[e.g.,][]{2016JCAP...12..012W,2017A&A...607A..18L, 2018ApJ...858....9L}. 
These are outflows launched from accretion disks around supermassive black holes \citep[SMBHs, e.g.,][]{2000ApJ...543..686P, King2015ARA&A..53..115K}. 
About 40\% of nearby AGNs host ultrafast outflows  \citep[UFOs, e.g.,][]{2010A&A...521A..57T, 2015MNRAS.451.4169G}, whose physical origins are thought to be AGN disk winds typically launched at a radius of $\sim100r_\mathrm{g}$ with wind velocities $v_\mathrm{w}\sim0.03$--$0.3c$ \citep[e.g.,][]{2015MNRAS.451.4169G} and at most $\sim0.6c$ \citep{2021ApJ...920...24C}, where $r_\mathrm{g}=GM_\mathrm{BH}/c^2$ is a gravitational radius, $c$ is the speed of light, and $M_\mathrm{BH}$ is an SMBH mass. 
Such winds can carry large kinetic power exceeding a few percent of their bolometric luminosity \citep{2015MNRAS.451.4169G, 2019ApJ...871..156M}.  

Several studies have proposed that shocks produced by the interaction of disk winds with the interstellar medium (ISM) of their host galaxy can accelerate high-energy cosmic rays (CRs), producing both gamma rays and neutrinos. 
In this framework, estimates of the AGN disk-wind contribution to the CNB span between $\sim 10\%$ up to nearly 100\%  \citep{2016JCAP...12..012W, 2017A&A...607A..18L, 2018ApJ...858....9L}.  
This broad range mainly reflects poorly constrained model parameters. 
For example, they assume very high gas densities \citep[e.g., $\sim10^3~\mathrm{cm^{-3}}$ at 100~pc in][]{2016JCAP...12..012W}, far above a typical ISM value of $\sim10~\mathrm{cm^{-3}}$ \citep[e.g.,][]{2025arXiv250401410S}. 
Also, they adopt low wind kinetic efficiency of only $\sim1$--5\% of the AGN luminosity, while recent results from the X-Ray Imaging and Spectroscopy Mission \citep[XRISM,][]{2020SPIE11444E..22T} report that the efficiency can reach $\sim100\%$ \citep[e.g.,][]{2025Natur.641.1132X, 2025ApJ...988L..54X}. 

\citet{Ajello2021ApJ...921..144A} reported GeV gamma-ray emission from Seyfert galaxies associated with disk winds using a stacking analysis of the \textit{Fermi} Large Area Telescope (LAT) survey data. 
In addition, the latest \textit{Fermi}-LAT source catalog \citep[4FGL-DR4,][]{2023arXiv230712546B} has reported detections of several sources categorized as Seyfert galaxies. 
Models based on wind-ISM shocks reproduce the observed GeV signals in nearby Seyfert galaxies, including NGC~1068 \citep{2016A&A...596A..68L,2022arXiv220702097I,2023MNRAS.526..181P}, NGC~4151 \citep{2025JCAP...07..013P}, and GRS~1734--292 \citep{2025ApJ...980..131S}. 
These measurements provide a path to calibrate key wind properties, such as their kinetic power and the ambient gas density.

In this work, we revisit the contribution of AGN disk winds to both CGB and CNB, calibrating model parameters against the latest GeV observations of nearby Seyfert galaxies. 
Section~\ref{sec:method} describes our wind dynamical evolution and emission model, multi-messenger spectral energy distributions (SEDs), and their parameter dependencies. 
A sample selection and parameter calibration to the gamma-ray fluxes from Seyfert galaxies are presented in Section~\ref{sec:sample}. 
Section~\ref{sec:CBR} presents contributions of disk winds to the CNB and CGB. 
We then discuss the origin of the gamma-ray emission in the individual Seyfert galaxies, compare our results with previous works, and outline prospects for future observations in Section~\ref{sec:discussion}. We present our conclusions and provide a summary of this work in Section~\ref{sec:conclusions}. 
Throughout this paper, we adopt a standard cosmology with ($h, \Omega_\mathrm{M}, \Omega_\Lambda$) = (0.7, 0.3, 0.7).  

\section{Disk Wind Model}\label{sec:method}

    We consider an individual AGN disk wind with a steady-state CR distribution.
    For analytical tractability, we model the wind as a spherically averaged system when calculating its dynamical evolution and non-thermal emission. 
    This approximation provides an effective one-dimensional description of the wind-driven forward and reverse shocks, and allows us to compute the resulting photon and neutrino luminosities for individual sources. 
    A solid-angle correction factor will later be applied (see Section~\ref{sec:CBR}) to account for the anisotropy of real disk winds. 
    
\subsection{Dynamical Evolution of an AGN Disk Wind}\label{sec:wind dynamics}

    A disk wind with a mass outflow rate of $\dot{M}_\mathrm{w}$ and a velocity of $v_\mathrm{w}$ is launched from an accretion disk around an SMBH with bolometric luminosity of $L_\mathrm{AGN}$. We adopt $v_\mathrm{w} = 0.1c$, which is a typical UFO velocity  \citep[e.g.,][]{2010A&A...521A..57T}, as a fiducial value. 
    We relate the momentum flux of the disk wind to the radiative momentum flux of the AGN by  
    \begin{align}\label{eq:pdot_wind}
    \dot{M}_\mathrm{w} v_\mathrm{w} = b_\mathrm{w} \frac{L_\mathrm{AGN}}{c},
    \end{align}
    where $b_\mathrm{w}$ is a dimensionless boost factor, which determines the kinetic momentum of winds \citep{2012MNRAS.425..605F}. 
    A choice of $b_\mathrm{w}=1$ corresponds to a momentum-conserving wind, while $b_\mathrm{w}=20$ corresponds to an energy-conserving system for $v_\mathrm{w}=0.1c$, where the kinetic power of the disk wind $\dot{K}_\mathrm{w}=\dot{M}_\mathrm{w}v_\mathrm{w}^2/2$ can be expressed as 
    \begin{align}
        \dot{K}_\mathrm{w}=\left(\frac{b_\mathrm{w}}{20}\right)\left(\frac{v_\mathrm{w}}{0.1c}\right)L_\mathrm{AGN}.\label{eq:Kdot_wind}
    \end{align}
    X-ray observations have indicated that 
    $b_\mathrm{w}$ values typically range from $10^{-2}$ to $1$ in most systems \citep{2015MNRAS.451.4169G}, although recent studies, including XRISM observations, reported higher values of $b_\mathrm{w}\sim10$--$10^3$~\citep[e.g.,][]{2019ApJ...871..156M, 2025Natur.641.1132X, 2025ApJ...988L..54X}. 
    As a characteristic value, we therefore set $b_\mathrm{w}=1$, unless otherwise noted. 

    When a disk wind interacts with the ISM of its host galaxy, it drives a forward shock into the ISM and a reverse shock into the wind.  
    The forward shock compresses the ISM to form the shocked ambient medium (SAM), while the reverse shock compresses the wind to create the shocked wind (SW); these regions are separated by a contact discontinuity.  
    
    We model the radial profile of ISM hydrogen gas density as a power-law, as suggested by observations of the Milky Way halo \citep[e.g.,][]{2015ApJ...800...14M, 2018ApJ...862...34N}:
    \begin{align}\label{eq:ISM profile}
        n_\mathrm{ISM}(r) = n_0 \left( \frac{r}{r_0} \right)^{-\beta},
    \end{align}
    where $n_0$ is the ISM gas density at radius $r_0$ from the SMBH. 
    While \citet[][]{2015ApJ...800...14M} find $\beta\sim1.5$, we adopt $\beta=1$ as a simple fiducial choice, following previous work on the dynamics of an AGN wind \citep{2012MNRAS.425..605F}, and explore the dependence of wind emission on $\beta$ explicitly in Section~\ref{sec:beta}. 
    We fix $r_0=100~\mathrm{pc}$, following \citet{2012MNRAS.425..605F, 2015MNRAS.447.3612N, 2024ApJ...968..116Y}.
    We set $n_0 = 10~\mathrm{cm}^{-3}$ as a fiducial value, consistent with typical ISM densities at $\sim100$~pc in AGN host galaxies \citep[e.g.,][]{2025arXiv250401410S}. 

    We model shock propagation following \citet{1992ApJ...388..103K}.
    For $\beta\leq11/7$, the SAM and the SW expand adiabatically, and the shock radii follow a self-similar solution. 
    We first consider the free-expansion phase ($t_\mathrm{w} < \tau_\mathrm{free}$), where $t_\mathrm{w}$ is the wind age and $\tau_\mathrm{free}$ is the free-expansion time defined by\footnote{This definition differs by a factor of 3 compared to \citet{2025ApJ...980..131S} because we instead adopt the fiducial time scale in \citet{1992ApJ...388..103K} as $\tau_\mathrm{free}$.}
    \begin{align}
    \tau_\mathrm{free} = \frac{r_0}{v_\mathrm{w}}
    \left[ \frac{(3-\beta)\dot{K}_\mathrm{w}}{6\pi \mu_\mathrm{H} m_p n_0 r_0^2 v_\mathrm{w}^3} \right]^{-1/(2-\beta)},
    \label{eq:fiducial time}
    \end{align}
    at which the mass of the ejected wind material becomes comparable to that of the swept-up medium.
    Here, $\mu_\mathrm{H}=1.4$ is the mean molecular weight and $m_p$ is the mass of a proton. 
    During this stage, the radii of the forward shock (FS), contact discontinuity (CD), and reverse shock (RS) are
    \begin{align}
    R_\mathrm{FS}(t_\mathrm{w}) &= \frac{v_\mathrm{w} t_\mathrm{w}}{\lambda_\mathrm{CD, free}}\label{eq:FS radius free}, \\
    R_\mathrm{CD}(t_\mathrm{w}) &= R_\mathrm{RS}(t_\mathrm{w}) = v_\mathrm{w} t_\mathrm{w}\label{eq:CDRS radii free}, 
    \end{align}
    where $\lambda_\mathrm{CD, free}\sim (215 - 54\beta)/(235 - 54\beta)$ is the ratio of the contact discontinuity radius to the forward shock radius in the free-expansion phase.
    In the adiabatic phase ($t_\mathrm{w} > \tau_\mathrm{free}$), the forward shock radius is
    \begin{align}
    R_\mathrm{FS}(t_\mathrm{w}) \simeq A_\mathrm{FS} 
    \left( \frac{\dot{K}_\mathrm{w}}{\mu_\mathrm{H} m_p n_0 r_0^\beta} \right)^{1/(5-\beta)} 
    t_\mathrm{w}^{3/(5-\beta)}\label{eq:FS radius},
    \end{align}
    where $A_\mathrm{FS}$ is a dimensionless factor
    \begin{align}
    A_\mathrm{FS} = \frac{2(3-\beta)(5-\beta)^3}{9\pi(11-\beta)(7-2\beta)\lambda_\mathrm{CD, adi}^3}
    \end{align}
    and
    $\lambda_\mathrm{CD, adi}\sim(123 - 8\beta)/(143 - 8\beta)$ is the ratio of the contact discontinuity radius to the forward shock radius in the adiabatic phase. 
    Therefore, the contact discontinuity radius is 
    \begin{align}
        R_\mathrm{CD}(t_\mathrm{w})\simeq \lambda_\mathrm{CD, adi}A_\mathrm{FS}\left( \frac{\dot{K}_\mathrm{w}}{\mu_\mathrm{H} m_p n_0 r_0^\beta} \right)^{1/(5-\beta)}t_\mathrm{w}^{3/(5-\beta)}\label{eq:CD radius}.
    \end{align}
    The reverse shock radius is
    \begin{align}
    R_\mathrm{RS}(t_\mathrm{w}) \simeq R_\mathrm{FS} \left[ \frac{1}{\lambda_\mathrm{CD,free}^3} + \frac{A_\mathrm{RS}}{\lambda_\mathrm{CD,adi}^3} \left( \frac{R_\mathrm{FS}}{v_\mathrm{w} \tau_\mathrm{free}} \right)^{(2-\beta)/2} \right]^{-1/3}\label{eq:RS radius},
    \end{align}
    with
    \begin{align}
    A_\mathrm{RS} = \left( \frac{15}{16} \right)^{15/4} \frac{2(7 - 2\beta)^{1/2}}{3(11 - \beta)}.
    \end{align}

\subsection{Particle Evolution in AGN Disk Wind Shocks}

    The forward shock and reverse shock accelerate electrons and protons to relativistic energies via diffusive shock acceleration \citep[DSA, e.g.,][]{1978MNRAS.182..147B, 1978MNRAS.182..443B, 1978ApJ...221L..29B, 1983RPPh...46..973D}.  
    Under the DSA mechanism, the particle acceleration rate is approximated by
    \begin{align}
    \tau_\mathrm{acc}^{-1} = \frac{3 e B_\mathrm{up} v_\mathrm{up}^2}{8 \eta_\mathrm{g} c E_\mathrm{CR}},
    \end{align}
    where $\eta_\mathrm{g}$ is the gyrofactor, $e$ is the elementary charge, $B_\mathrm{up}$ is the magnetic field in the upstream region (i.e., the ISM for the forward shock or the disk wind for the reverse shock), and $v_\mathrm{up}$ is the upstream velocity in the shock rest frame \citep{1983RPPh...46..973D}. $\eta_\mathrm{g} = 1$ corresponds to the Bohm limit, which we adopt as our fiducial choice \citep{2006NatPh...2..614S}. 
    We parameterize the SAM magnetic field as $B_\mathrm{SAM}$, while in the SW it is defined through the magnetic energy fraction,   
    \begin{align}\label{eq:epsilon_B,SW def}
    U_{B, \mathrm{SW}}=\frac{\epsilon_{B, \mathrm{SW}}\dot{K}_\mathrm{w}}{4\pi R_\mathrm{RS}^2v_\mathrm{w}},  
    \end{align}
    where $U_{B, \mathrm{SW}}$ is the energy density of the magnetic field in the SW region. 
    We set $B_\mathrm{SAM} = 100\ \mathrm{\mu G}$ as the fiducial value, slightly above the typical ISM field of $\sim 10\ \mathrm{\mu G}$ \citep[e.g.,][]{2015A&ARv..24....4B}, motivated by magnetic field amplification at shocks \citep{2013MNRAS.431..415B, 2025arXiv251013946N}. 
    The SW magnetic energy fraction $\epsilon_{B,\mathrm{SW}}$ is set to $10^{-5}$ (see Section~\ref{sec:B_SW} for details). 
    $v_\mathrm{up}$ is written as $V_\mathrm{FS}$ at the forward shock and $v_\mathrm{w}-V_\mathrm{RS}$ at the reverse shock, respectively, where $V_\mathrm{FS/RS}$ is the velocity of the forward shock/reverse shock in the SMBH rest frame.  
    These parameters determine the maximum energy CRs can attain in our system, as they establish the balance between acceleration, advection and energy loss processes. 
    
    After acceleration, CRs reside in downstream regions (i.e., the SAM or SW), where they lose energy through several processes.  
    For electrons, the dominant cooling mechanisms are synchrotron radiation and inverse Compton (IC) scattering \citep{1970RvMP...42..237B}.  
    The synchrotron cooling rate is \citep{1986rpa..book.....R}
    \begin{align}
    \tau_\mathrm{sync}^{-1}
    &= \frac{4 c \sigma_\mathrm{T}}{3 m_e c^2} \gamma_e U_B, 
    \end{align}
    where $\sigma_\mathrm{T}$ is the Thomson cross section, $\gamma_e$ is the electron Lorentz factor, and $U_B$ is the magnetic energy density.
    The IC cooling rate is \citep{1986rpa..book.....R}
    \begin{align}
    \tau_\mathrm{IC}^{-1} &= \frac{4 c \sigma_\mathrm{T}}{3 m_e c^2} 
    \gamma_e U_\mathrm{ph} F_\mathrm{KN}(\gamma_e),
    \end{align}
    where $U_\mathrm{ph}$ is the photon energy density, and $F_\mathrm{KN}(\gamma_e)$ accounts for Klein-Nishina suppression:
    \begin{align}
    F_\mathrm{KN}(\gamma_e) = \frac{1}{U_\mathrm{ph}}
    \int_0^\infty d\epsilon_\mathrm{ph} 
    \frac{dn_\mathrm{ph}}{d\epsilon_\mathrm{ph}}\epsilon_\mathrm{ph} 
    f_\mathrm{KN} \left( \frac{4 \gamma_e \epsilon_\mathrm{ph}}{m_e c^2} \right),
    \end{align}
    where $\epsilon_\mathrm{ph}$ is the target photon energy, $dn_\mathrm{ph}/d\epsilon_\mathrm{ph}$ is the differential photon number density, and the function $f_\mathrm{KN}$ accounts for the suppression of the 
    Compton scattering cross-section in the  Klein-Nishina regime  \citep[see][for details]{2005MNRAS.363..954M}. 
    Here, target photon fields include external radiation from the accretion disk, dusty torus, and the cosmic microwave background. 
    \begin{figure*}
        \centering
        \includegraphics[width=0.95\linewidth]{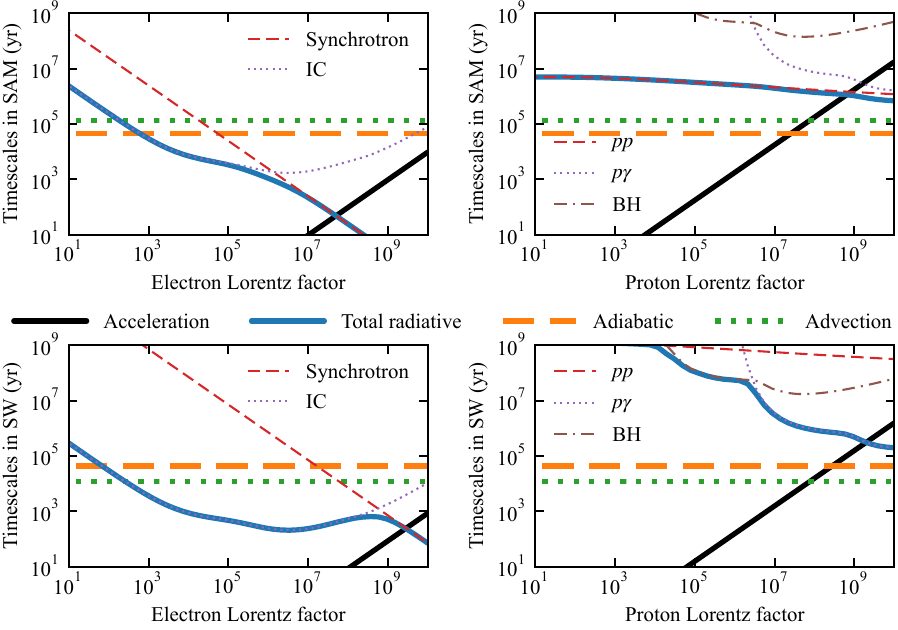}
        \caption{
        Timescales for CR acceleration, advection, and energy losses in the SAM and SW regions, computed using the fiducial parameters listed in Table~\ref{tab:fiducial parameters}.
        The top-left, top-right, bottom-left, and bottom-right panels show the timescales for CR electrons in the SAM, CR protons in the SAM, CR electrons in the SW, and CR protons in the SW, respectively.
        The black solid, blue solid, orange dashed, and green dotted lines represent the timescales for acceleration, total radiative losses, adiabatic losses, and advection, respectively.
        For electrons (left panels), the total radiative loss consists of synchrotron (red dashed) and inverse Compton (purple dotted) cooling.
        For protons (right panels), the total radiative loss includes \textit{pp} interactions (red dashed), $p\gamma$ interactions (purple dotted), and BH pair production (brown dash-dotted).
        }
        \label{fig:timescales}
\end{figure*}
\begin{table*}
    \centering
    \caption{Fiducial values of model parameters}
    \begin{tabular}{lcccc}
        \hline
        \textbf{Quantity} & \textbf{Parameter} & \textbf{Value} & \textbf{Unit} & \textbf{Reference} \\
        \hline
        \multicolumn{5}{c}{\textit{Literature value}}\\
        \hline 
        AGN bolometric luminosity & $L_\mathrm{AGN}$ & $10^{46}$ & erg s$^{-1}$ & \cite{2012MNRAS.425..605F}\\
        Wind velocity & $v_\mathrm{w}$ & 0.1 & $c$ & \cite{tombesi_unification_2013} \\
        Wind boost factor & $b_\mathrm{w}$ & 1 & -- & \cite{2015MNRAS.451.4169G} \\
        Wind age & $t_\mathrm{w}$ & $10^{5}$ & yr & \cite{2015MNRAS.451.2517S} \\
        ISM proton density at 100~pc & $n_0$ & 10 & cm$^{-3}$ & \cite{2025arXiv250401410S} \\
        Power-law index of ISM spatial distribution & $\beta$ & 1 & -- & \cite{2012MNRAS.425..605F}\\
        Temperature of AGN accretion disk & $k_\mathrm{B}T_\mathrm{disk}$ & 10 & eV & \cite{2008bhad.book.....K}\\
        Temperature of AGN dusty torus & $k_\mathrm{B}T_\mathrm{torus}$ & 0.03 & eV & \cite{2015ARAA..53..365N}\\
        Magnetic field strength in SAM region & $B_\mathrm{SAM}$ & 100 & $\mu$G & \cite{2013MNRAS.431..415B}\\
         & & & & \cite{2025arXiv251013946N}\\
        Magnetic-to-kinetic energy fraction in the SW & $\epsilon_{B, \mathrm{SW}}$ & $10^{-5}$ & -- & Section~(\ref{sec:B_SW})\\
        Power-law index of injected CRs & $q_\mathrm{CR}$ & 2.0 & -- & \cite{1983RPPh...46..973D}\\
        Energy fraction of CR electrons to thermal energy & $\xi_e$ & 0.01 & -- & \cite{2013Sci...339..807A} \\
        Energy fraction of CR protons to thermal energy & $\xi_p$ & 0.1 & -- & \cite{2013Sci...339..807A}\\
         & & & & \cite{2015PhRvL.114h5003P}\\
        Gyrofactor & $\eta_\mathrm{g}$ & 1 & -- & \cite{2006NatPh...2..614S}\\
        \hline
        \multicolumn{5}{c}{\textit{Derived value}}\\
        \hline 
        Kinetic power of disk wind & $\dot{K}_\mathrm{w}$ & $5\times10^{44}$ & erg s$^{-1}$ & Equation~(\ref{eq:Kdot_wind})\\
        Free-expansion timescale & $\tau_\mathrm{free}$ & 2.9 & yr & Equation~(\ref{eq:fiducial time})\\
        Forward shock radius & $R_\mathrm{FS}$ & 290 & pc & Equations~(\ref{eq:FS radius free}, \ref{eq:FS radius})\\
        Reverse shock radius & $R_\mathrm{RS}$ & 102 & pc & Equations~(\ref{eq:CDRS radii free}, \ref{eq:RS radius})\\
        Magnetic field in the SW region & $B_\mathrm{SW}$ & 5.8 & $\mu$G & Equation~(\ref{eq:epsilon_B,SW def})\\
        \hline
    \end{tabular}
    \\
    \raggedright
    \textit{Top}: Fiducial values of model parameters based on literature.
    \textit{Bottom}: Values of quantities derived from model parameters in the top panel and equations whose reference numbers are shown in the right column.
    \label{tab:fiducial parameters}
\end{table*}
    The disk emission is modeled as an isotropic multi-color blackbody with an inner temperature of $T_\mathrm{disk}$, normalized via the bolometric-ultraviolet (UV) luminosity correlation at $1400~\mathrm{\AA}$ \citep{2019MNRAS.488.5185N}.  
    The torus emission is modeled as a single-temperature blackbody with $T_\mathrm{torus}$, normalized via the hard X-ray--mid-IR luminosity correlation at $12\ \mathrm{\mu m}$ \citep{2017ApJ...835...74I}. As fiducial choices, the disk and torus temperatures are set to $k_\mathrm{B}T_\mathrm{disk} = 10\ \mathrm{eV}$ \citep[e.g.,][]{1999PASP..111....1K, 2008bhad.book.....K} and $k_\mathrm{B}T_\mathrm{torus} = 0.03\ \mathrm{eV}$ \citep[e.g.,][]{2015ARAA..53..365N}, respectively, where $k_\mathrm{B}$ is the Boltzmann constant. 
    
    For protons, we consider energy losses via \textit{pp} interactions \citep{2006PhRvD..74c4018K}, $p\gamma$ interactions \citep{2008PhRvD..78c4013K}, and Bethe-Heitler (BH) pair production \citep{2008PhRvD..78c4013K}.  
    The \textit{pp} interaction rate is 
    \begin{align}
    \tau_{pp}^{-1}
    &= \kappa_{pp} \sigma_{pp} c n_p, \label{eq:t_pp} 
    \end{align}
    where $\kappa_{pp} \approx 0.5$ is the inelasticity, $\sigma_{pp}$ is the cross section, and $n_p$ is the target proton density \citep{2006PhRvD..74c4018K} considering the Rankine-Hugoniot jump condition.
    We set $n_p=4n_\mathrm{ISM}(R_\mathrm{FS})$ in the SAM and $n_p=4\dot{M}_\mathrm{w}/[4\pi R_\mathrm{RS}^2(v_\mathrm{w}-V_\mathrm{RS})\mu_\mathrm{H}m_p]$ in the SW.
    We use the \textit{pp} interaction cross section given by \citet{2006PhRvD..74c4018K} in our calculations. 
    The $p\gamma$ interaction loss rate is
    \begin{align}
    \tau_{p\gamma}^{-1} = \frac{c}{2 \gamma_p}
    \int_0^\infty &d\epsilon_\mathrm{ph} \frac{dn_\mathrm{ph}}{d\epsilon_\mathrm{ph}} \frac{1}{\epsilon_\mathrm{ph}^2}\\
    &\times\int_0^{2\gamma_p \epsilon_\mathrm{ph}} d\epsilon_\mathrm{ph}' \epsilon_\mathrm{ph}' 
    \sigma_{\gamma p}(\epsilon_\mathrm{ph}') K_{\gamma p}(\epsilon_\mathrm{ph}'),\nonumber
    \end{align}
    where $\gamma_p$ is the proton Lorentz factor, $\epsilon_\mathrm{ph}'$ is the energy of a target photon in the proton rest frame, $\sigma_{\gamma p}$ is the interaction cross section and $K_{\gamma p}$ is the inelasticity \citep[see][for detail]{2009herb.book.....D}.
    The BH pair production loss rate is 
    \begin{align}
    \tau_\mathrm{BH}^{-1} &= \frac{7 \alpha_\mathrm{f} (m_e c^2)^3 \sigma_\mathrm{T}}{9 \sqrt{2} \pi m_p c \gamma_p^2}
    \int_{\gamma_p^{-1} m_e c^2}^\infty d\epsilon_\mathrm{ph} \frac{1}{\epsilon_\mathrm{ph}^2} \frac{dn_\mathrm{ph}}{d\epsilon_\mathrm{ph}}(\epsilon_\mathrm{ph}) \nonumber\\
    &\quad \times \left\{ \left( \frac{2\gamma_p \epsilon_\mathrm{ph}}{m_e c^2} \right)^{3/2} \left[ \ln\left( \frac{2\gamma_p \epsilon_\mathrm{ph}}{m_e c^2} \right) - \frac{2}{3} \right] + \frac{2}{3} \right\},\nonumber
    \end{align}
    where $\alpha_\mathrm{f}$ is the fine-structure constant \citep{2009herb.book.....D}.

    We also account for adiabatic losses, which arise at a rate of 
    \begin{align}
    \tau_\mathrm{adi}^{-1} = \frac{1}{3} \frac{d\ln \mathcal{V}}{dt_\mathrm{w}},
    \end{align}
    where $\mathcal{V}$ is the volume of the shocked region  
    $\mathcal{V} = 4\pi(R_\mathrm{FS}^3 - R_\mathrm{CD}^3)/3$ for the SAM, and $\mathcal{V} = 4\pi(R_\mathrm{CD}^3 - R_\mathrm{RS}^3)/3$ for the SW \citep{2009herb.book.....D}.
    Finally, a fraction of CRs escape via advection.  
    We set the advection rates to $\tau_\mathrm{adv}^{-1} = V_\mathrm{FS}/R_\mathrm{FS}$ for the SAM and $\tau_\mathrm{adv}^{-1} = (v_\mathrm{w}/4 - V_\mathrm{RS})/R_\mathrm{FS}$ for the SW.
    
    Figure~\ref{fig:timescales} shows the acceleration, cooling, and advection timescales for our fiducial model parameters, listed in Table~\ref{tab:fiducial parameters}.  
    We adopt $t_\mathrm{w} = 10^5\ \mathrm{yr}$ \citep[the typical timescale for a cycle of AGN activity,][]{2015MNRAS.451.2517S}.    
    In both the SAM and SW, electrons in the Thomson regime cool mainly via IC scattering, with synchrotron losses becoming more important at higher energies due to the Klein-Nishina effect.
    For protons, adiabatic losses dominate over the entire energy range, with advection slightly more efficient than adiabatic expansion in the SAM.

\begin{figure*}[t]
    \centering
    \includegraphics[width=0.8\linewidth]{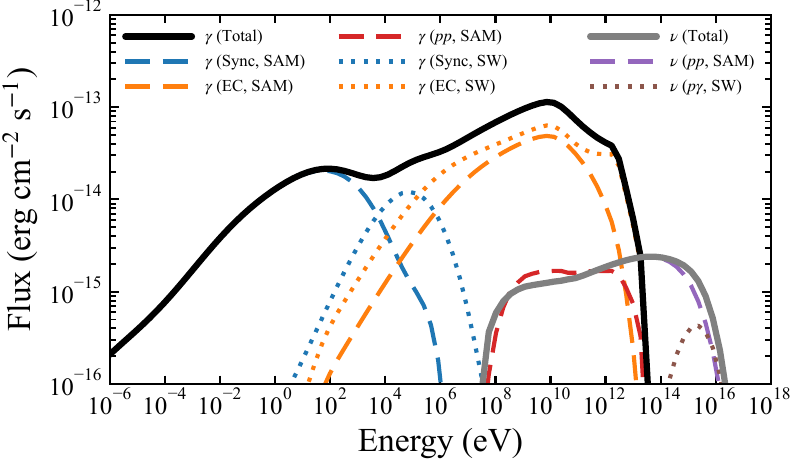}
    \caption{
    SEDs of the multiwavelength photon and neutrino emission from CRs accelerated by an AGN disk wind. 
    Calculations adopt a redshift of $z=0.02$ and the fiducial parameters in Table~\ref{tab:fiducial parameters}.
    Dashed and dotted lines show emission from the SAM and the SW, respectively. 
    Blue, orange, and red lines represent photon emission from synchrotron, EC, and \textit{pp} interactions. 
    Purple and brown lines show the sum of neutrino and antineutrino (per flavor) emission from \textit{pp} interactions and $p\gamma$ interactions, respectively.
    }
    \label{fig:fiducial SEDs}
\end{figure*}

    We calculate the steady-state CR spectrum by solving the transport equation
    \begin{align}\label{eq:transport}
    \frac{d}{dE_\mathrm{CR}}\left( \frac{E_\mathrm{CR}}{\tau_\mathrm{loss}} \frac{dN_\mathrm{CR}}{dE_\mathrm{CR}} \right) 
    + \tau_\mathrm{adv}^{-1} \frac{dN_\mathrm{CR}}{dE_\mathrm{CR}} 
    = \frac{d\dot{N}_\mathrm{CR}^{\mathrm{(inj)}}}{dE_\mathrm{CR}},
    \end{align}
    where $\tau_\mathrm{loss}^{-1}$ is the total energy-loss rate of CRs.  
    For electrons, $\tau^{-1}_\mathrm{loss} = \tau^{-1}_\mathrm{sync} + \tau^{-1}_\mathrm{IC} + \tau^{-1}_\mathrm{adi}$,  
    and for protons, $\tau^{-1}_\mathrm{loss} = \tau^{-1}_{pp} + \tau^{-1}_{p\gamma} + \tau^{-1}_\mathrm{BH} + \tau^{-1}_\mathrm{adi}$.  
    $d\dot{N}_\mathrm{CR}^{\mathrm{(inj)}}/dE_\mathrm{CR}$ is the injection rate of CRs with energy $E_\mathrm{CR}$.
    We consider that the system remains approximately steady over an observational timescale ($\sim 10$ yr).  
    This is justified because the shock and cooling properties do not significantly evolve over such timescales, and both are much shorter than both the AGN-phase lifetime or cooling timescales of CRs over most of the energy range we consider (see Figure~\ref{fig:timescales}).  
    Equation~(\ref{eq:transport}) is solved following \citet{1966SvPhU...9..223G}. 
    The injected CR spectrum is modeled as an exponential cutoff power law, 
    \begin{align}
    \frac{d\dot{N}_\mathrm{CR}^{\mathrm{(inj)}}}{dE_\mathrm{CR}} \propto E_\mathrm{CR}^{-q_\mathrm{CR}} \exp\left(-\frac{E_\mathrm{CR}}{E_{\mathrm{CR,max}}}\right),
    \end{align}
    with the same index of $q_\mathrm{CR}$ for electrons and protons. We set $q_\mathrm{CR} = 2.0$ as a fiducial value, consistent with standard DSA theory \citep[e.g.,][]{1983RPPh...46..973D}.
    The maximum energy $E_{\mathrm{CR,max}}$ is determined by the condition
    \begin{align}
    \tau_\mathrm{acc}^{-1}(E_{\mathrm{CR,max}}) = \tau_\mathrm{adv}^{-1} + \tau_\mathrm{loss}^{-1}(E_{\mathrm{CR,max}}).
    \end{align} 
    The CR injection rate is normalized such that a fraction $\xi_\mathrm{CR}(\in\{\xi_e, \xi_p\})$ of the thermal energy injection rate into the shocked region is converted to CRs. 
    We adopt $\xi_p = 0.1$ for protons based on supernova remnant (SNR) observations \citep[e.g.,][]{2013Sci...339..807A}. 
    This is because particle acceleration mechanisms in SNRs are thought to be DSA, and the ranges of SNR parameters (e.g., shock velocities and ambient magnetic field) are similar to those of disk wind parameters.
    For electrons, we adopt $\xi_e = 0.01$ based on Particle-in-Cell simulations \citep[e.g.,][]{2015PhRvL.114h5003P}.
    For the SAM, the thermal energy injection rate is \citep{1992ApJ...388..103K}  
    \begin{align}\label{eq:Qdot_SAM}
    \dot{Q}_\mathrm{SAM} \simeq \frac{9(5-\beta)(143 - 8\beta)^3}{4(11-\beta)(7-2\beta)(123 - 8\beta)^3} \ \dot{K}_\mathrm{w},
    \end{align}
    and for the SW, it is
    \begin{align}\label{eq:Qdot_SW}
    \dot{Q}_\mathrm{SW} \simeq \frac{5-\beta}{11-\beta} \ \dot{K}_\mathrm{w}.
    \end{align}
    The normalization condition for CR injection in the SAM/SW is therefore
    \begin{align}
    \xi_\mathrm{CR} \ \dot{Q}_\mathrm{SAM/SW} = \int_{m_\mathrm{CR}c^2}^\infty dE_\mathrm{CR} \ \frac{d\dot{N}_\mathrm{CR}^{\mathrm{(inj)}}}{dE_\mathrm{CR}} \ E_\mathrm{CR},
    \end{align}
    where $m_\mathrm{CR}$ is the rest mass of a CR.

    \subsection{Photon and Neutrino Emission and Propagation}
    
    High-energy CRs produce photons, neutrinos, and other secondary particles through several channels.  
    In our model, we include synchrotron radiation \citep{1986rpa..book.....R}, IC scattering \citep{2014ApJ...783..100K}, \textit{pp} interactions \citep{2006PhRvD..74c4018K}, $p\gamma$ interactions \citep{2008PhRvD..78c4013K}, and BH pair production \citep{1970RvMP...42..237B, 2008PhRvD..78c4013K}.  
    Among these, \textit{pp} interactions, $p\gamma$ interactions, and BH processes also generate secondary electrons and positrons.  
    We compute their resulting synchrotron and IC emission self-consistently.
    For IC scattering, we include both synchrotron self-Compton and external Compton (EC) components.  
    The AGN accretion disk, dusty torus, and the cosmic microwave background provide the seed photons for EC.  
    We account for the absorption of gamma rays by pair-production in these soft radiation fields ($\gamma + \gamma \rightarrow e^+ + e^-$),  
    and additionally consider the effect of $\gamma\gamma$ attenuation as photons propagate cosmologically through the 
   extragalactic background light (EBL), using the EBL model of \citet{2022ApJ...941...33F}.
   We do not include cascade secondary emission.
   Then, the observed photon or neutrino flux from a disk wind at a redshift of $z$ is given as
   \begin{align}
       E^2\frac{dN}{dE}(E)&=\frac{L'(E')}{4\pi d_\mathrm{L}(z)^2}\exp{[-\tau_\mathrm{EBL}(z, E)]},
   \end{align}
   where $L'(E')=E'^2d^2N/(dt'dE')$ is the differential luminosity at energy $E'=(1+z)E$ in the source rest-frame, and $d_\mathrm{L}(z)$ is the luminosity distance.
   The EBL $\gamma\gamma$ opacity $\tau_\mathrm{EBL}(z,E)$ is taken into account in our calculations for gamma rays above energies of $E\geq10$~GeV~\citep{2022ApJ...941...33F}. 
    
\begin{figure*}[th!]
    \centering
    \includegraphics[width=0.8\linewidth]{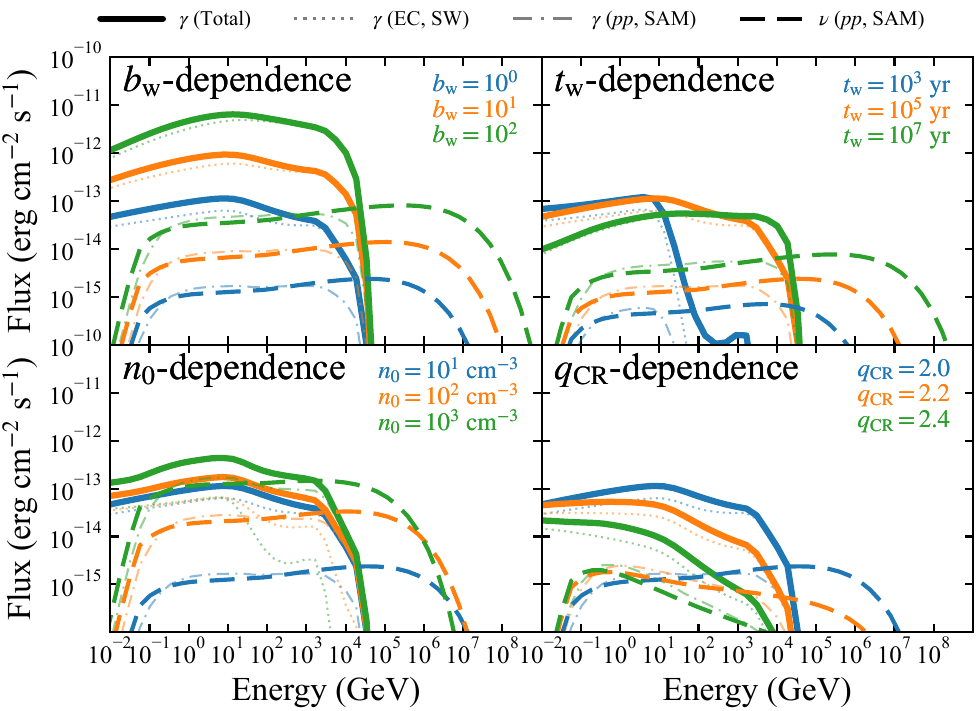}
    \caption{
    Parameter dependence of the SEDs. 
    In all panels, 
    thick solid lines, thin dotted lines, thin dashed-dotted lines, and thick dashed lines represent total gamma rays, gamma rays from the SW via EC scattering, those from the SAM via \textit{pp} interactions, and the sum of neutrinos and antineutrinos per flavor from the SAM via \textit{pp} interactions, respectively. 
    \textit{Top-Left}: Dependence on the boost factor $b_\mathrm{w}$. Blue, orange, and green lines correspond to $b_\mathrm{w}=1$, 10, and 100, respectively. 
    \textit{Top-Right}: Dependence on the wind age $t_\mathrm{w}$. Blue, orange, and green lines correspond to $t_\mathrm{w}=10^3$~yr, $10^5$~yr, and $10^7$~yr. 
    \textit{Bottom-Left}: Dependence on the gas density $n_0$. Blue, orange, and green lines correspond to $n_0=10\ \mathrm{cm}^{-3}$, $10^2\ \mathrm{cm}^{-3}$, and $10^3\ \mathrm{cm}^{-3}$. 
    \textit{Bottom-Right}: Dependence on the CR index $q_\mathrm{CR}$. Blue, orange, and green lines correspond to $q_\mathrm{CR}=2.0$, $2.2$, and $2.4$. 
    In each panel, the other parameters are fixed to the fiducial values (i.e., $b_\mathrm{w}=1$, $t_\mathrm{w}=10^5$~yr, $n_0=10~\mathrm{cm}^{-3}$, and $q_\mathrm{CR}=2.0$).
    }
    \label{fig:parameter dependence}
\end{figure*}

\subsection{Lepto-hadronic SEDs of AGN Disk Winds}

Figure~\ref{fig:fiducial SEDs} shows the resulting multi-messenger SEDs for our disk wind model located at a redshift of $z=0.02$, using the fiducial parameter set in Table~\ref{tab:fiducial parameters}. Gamma-ray attenuation by both EBL and internal photons is included. For completeness, we also compute synchrotron self-Compton in both SAM and SW regions, \textit{pp} interactions in the SW, $p\gamma$ interactions in the SAM, and emission from secondary particles, but these components are subdominant and are not labeled in the plot.

Below the keV band, synchrotron emission from the SAM dominates. Its spectral peak at $\sim100$~eV reflects the maximum electron Lorentz factor in the SAM, $\gamma_{e}\approx4\times10^7$ (see the top-left panel of Figure~\ref{fig:timescales}). 
From a few keV to the sub-MeV band, synchrotron from the SW becomes dominant. The SW synchrotron spectrum is harder than the SAM synchrotron because Compton cooling is suppressed by the Klein-Nishina effect (see the bottom-left panel of Figure~\ref{fig:timescales}). The peak at $\sim100$~keV corresponds to the maximum Lorentz factor of primary electrons in the SW. 

Above the MeV band, EC emission from both regions dominates for fiducial parameters. Because the SW lies closer to the disk and the torus ($R_\mathrm{RS}=102~\mathrm{pc}$ while $R_\mathrm{FS}=290~\mathrm{pc}$), the SW EC component is slightly brighter. The sharp cutoff near $\sim10$~TeV arises from EBL absorption. Since synchrotron cooling is stronger at high energies and the magnetic field in the SAM ($100~\mathrm{\mu G}$) is stronger than that in the SW ($5.8~\mathrm{\mu G}$), the EC component in the SAM has a softer spectral shape above $\sim10$~GeV.

With the fiducial parameters, hadronic channels would make a modest contribution (see also Figure~\ref{fig:timescales}). Because the ISM density is higher than the gas density in the SW, $pp$ interactions in the SAM dominate the hadronic emission. Above a PeV, neutrinos from $p\gamma$ interactions in the SW become comparable to those from \textit{pp} interactions in the SAM. 

\begin{table*}
    \centering
    \caption{Observed properties of the \textit{Fermi}-detected Seyfert galaxies}
    \begin{tabular}{lccccc}
    \hline \hline
    Object & $d$  & $L_\gamma$ & $L_\mathrm{AGN}$ & $L_\mathrm{X}$ & $v_\mathrm{w}$ \\
    & (Mpc) & (erg\,s$^{-1}$) & (erg\,s$^{-1}$) & (erg\,s$^{-1}$) & ($c$)\\
    & (1) & (2) & (3) & (4) & (5)\\
    \hline
    Circinus & 4.21 & $1.2\times10^{40}$ & $1.5\times10^{44}$ & $1.2\times10^{42}$ & None \\
    NGC 4945 & 3.80 & $2.1\times10^{40}$ & $1.0\times10^{44}$ & $2.2\times10^{42}$ & None \\
    NGC 4151 & 15.1 & $3.0\times10^{40}$ & $1.4\times10^{44}$ & $1.5\times10^{43}$ & 0.105 \\
    NGC 1068 & 14.4 & $1.7\times10^{41}$ & $1.8\times10^{45}$ & $1.2\times10^{42}$ & 0.280 \\
    GRS 1734--292 & 93.2 & $1.2\times10^{43}$ & $3.8\times10^{45}$ & $1.2\times10^{44}$ & None \\
    \hline
    \end{tabular}
    \\
    \raggedright
    Column descriptions: (1) distance; (2) gamma-ray luminosity calculated from the 100~MeV--100~GeV flux \citep{2023arXiv230712546B}; (3) AGN bolometric luminosity; (4) hard X-ray luminosity integrated over 14--195~keV (converted from logarithmic values); (5) UFO velocity \citep[][for NGC~4151 and NGC~1068, respectively]{2010A&A...521A..57T, 2019ApJ...871..156M}. ``None'' indicates no UFO detection. Distance references are \citet[][Tully--Fisher]{2009AJ....138..323T} for Circinus, \citet[][Tip of the red-giant branch]{2016MNRAS.457.1419M} for NGC~4945, \citet[][Cepheid]{2020ApJ...902...26Y} for NGC~4151, and \citet[][Tully--Fisher]{1988cng..book.....T} for NGC~1068. 
    Distance to GRS~1734--292 is derived from its redshift $z=0.021$ \citep{https://doi.org/10.26132/ned1}.
    Bolometric luminosities are cited from \citet[][for the Circinus galaxy]{2017MNRAS.472.3854S}, \citet[][for NGC~4945]{2000A&A...357...24M}, \citet[][for NGC~4151]{2020MNRAS.493.3893K}, and \citet[][for NGC~1068]{2020A&A...634A...1G}. We derive the bolometric luminosity of GRS~1734--292 using Equation~(\ref{eq:bolometric_correlation}).
    \label{tab:Fermi_Seyfert galaxies}
\end{table*}

We further investigate how the SED depends on key model parameters. 
Figure~\ref{fig:parameter dependence} presents the total gamma-ray emission, the gamma-ray emission from EC scattering in the SW, and that from \textit{pp} interactions in the SAM, together with neutrino emission from \textit{pp} interactions in the SAM. 
We vary the wind boost factor $b_\mathrm{w}$ (top-left panel), the wind age $t_\mathrm{w}$ (top-right panel), the ISM gas density $n_0$ (bottom-left panel), and the injected CR spectral index $q_\mathrm{CR}$ (bottom-right panel), while keeping all other parameters fixed to their fiducial values.
First, by varying the wind boost factor $b_\mathrm{w}$ (top-left panel), we find that stronger winds produce higher gamma-ray and neutrino fluxes. Both the EC component from the SW and the $pp$ component from the SAM increase in normalization. Their spectral shapes and the dominance of EC over $pp$ gamma-rays remain almost unchanged. 
Second, by varying the wind age $t_\mathrm{w}$ (top-right panel), we find that older winds yield brighter hadronic emission from the SAM. The SW EC component changes weakly, although younger systems show stronger gamma-ray attenuation due to closer distance of the SW to the disk. Because of longer advection timescales, the cutoff energy in the resulting neutrino spectra shifts to higher energies for older winds. 
Third, varying the ISM density $n_0$ (bottom-left panel) shows that winds expanding into denser media produce more \textit{pp} interactions in the SAM. For the SW EC component, gamma-ray attenuation becomes stronger in denser environments while changes in the intrinsic luminosity are only modest. 
Finally, changing the injected CR index $q_\mathrm{CR}$ (bottom-right panel) primarily affects the overall spectral shape. 

These trends can be understood with simple scaling arguments. 
Electrons with $\gamma_e\sim10^4$--$10^6$ produce GeV--TeV gamma rays. In this energy regime, EC cooling dominates other losses (see the bottom-left panel of Figure~\ref{fig:timescales}), so the EC radiative efficiency is close to unity. Therefore, the EC luminosity roughly linearly scales with the wind power and, hence, with $b_\mathrm{w}$ for fixed $\xi_e$. 
For \textit{pp} interactions in the SAM, the \textit{pp} efficiency can be written as ${\tau_{pp}^{-1}}/{(\tau_{p,\mathrm{loss}}^{-1}+\tau_\mathrm{adv}^{-1})}\approx\tau_\mathrm{adv}/\tau_{pp}$, where advection and adiabatic losses, both $\propto t_\mathrm{w}$, are typically faster than other proton loss channels (see Figure~\ref{fig:timescales}). Using the $pp$ interaction timescale (Equation~\ref{eq:t_pp}), the \textit{pp} luminosity in the SAM can be approximated as 
\begin{align}
L'_{pp,\gamma/\nu,\mathrm{SAM}}
  &\simeq \epsilon_{\gamma/\nu}\xi_p \dot{Q}_\mathrm{SAM}\frac{\tau_\mathrm{adv}}{\tau_{pp}} \label{eq:pp luminosity}\\
  &\propto b_\mathrm{w}^{(5-2\beta)/(5-\beta)}\, t_\mathrm{w}^{(5-4\beta)/(5-\beta)}\, n_0^{5/(5-\beta)},\label{eq:pp dependence}
\end{align}
where the dimensionless factor $\epsilon_{\gamma/\nu}$ denotes the fraction of the dissipated CR power transferred to a given gamma-ray or neutrino channel. For the fiducial density slope $\beta=1$, this gives $L'_{pp,\gamma/\nu,\mathrm{SAM}}\propto b_\mathrm{w}^{3/4}\, t_\mathrm{w}^{1/4}\, n_0^{5/4}$, which reproduces the monotonic increase of the \textit{pp}-induced gamma-ray and neutrino fluxes with $b_\mathrm{w}$, $t_\mathrm{w}$, and $n_0$ seen in Figure~\ref{fig:parameter dependence}.

The maximum proton energy in the SAM also increases with wind age, 
$E_{p,\max}\propto t_\mathrm{w}^{(1+\beta)/(5-\beta)}$, 
set by the balance between acceleration and adiabatic losses (see the top-left panel of Figure~\ref{fig:timescales}). 
This explains the shift of the neutrino cutoff energy with $t_\mathrm{w}$ in the top-right panel. Similarly, in denser media, the forward shocks and reverse shocks are located at smaller radii with the fixed wind age to $t_\mathrm{w}=10^5$~yr; in particular $R_\mathrm{RS}\propto n_0^{-1/2(5-\beta)}$ for $t_\mathrm{w}>\tau_\mathrm{free}$ (from Equations~\ref{eq:fiducial time}, \ref{eq:FS radius}, and \ref{eq:RS radius}). 
As a result, the gamma rays traverse stronger UV/optical radiation fields from the disk and torus, which enhances $\gamma\gamma$ absorption and suppresses the EC spectrum at $\gtrsim10$~GeV for large $n_0$.
  
\begin{table*}
    \centering
    \caption{Model parameters for \textit{Fermi}-detected Seyfert galaxies in Model~A and Model~B scenarios}
    \begin{tabular}{llccccccc}
    \hline \hline
    Object & Model~& $b_\mathrm{w}$ & $B_\mathrm{SAM}$ & $t_\mathrm{w}$ & $n_0$ & $q_\mathrm{CR}$ & $R_\mathrm{FS}$ & $R_\mathrm{RS}$\\
           &        & -- & ($\mu$G) & (yr) & (cm$^{-3}$) & -- & (pc) & (pc)\\
           &        & (1) & (2) & (3) & (4) & (5) & (6) & (7)\\
    \hline
    Circinus & A & 8 & 450 & $5\times10^3$ & 10$^\dagger$ & 2.30 & 18 & 7 \\
             & B & 1 & 170 & $10^5$$^\dagger$ & 100 & 2.24$^\dagger$ & 57 & 9.1 \\
    NGC~4945 & A & 50 & 400 & $7\times10^3$ & 10 $^\dagger$ & 2.40 & 33 & 15 \\
             & B & 5 & 170 & $10^5$$^\dagger$ & 75 & 2.30$^\dagger$ & 83 & 16 \\
    NGC~4151 & A & 40 & 59 & $1\times10^5$ & 10$^\dagger$ & 2.27 & 254 & 82 \\
             & B & 3 & 230 & $10^5$$^\dagger$ & 100 & 2.11$^\dagger$ & 75 & 13 \\
    NGC~1068 & A & 2 & 2000 & 3000 & 10$^\dagger$ & 2.20 & 21 & 6.8 \\
             & B & 1 & 350 & $10^5$$^\dagger$ & 100 & 2.35$^\dagger$ & 137 & 20 \\
    GRS~1734--292 & A & 130 & 120 & $2\times10^3$ & 10$^\dagger$ & 2.30 & 41 & 30 \\
                  & B & 130 & 17 & $10^5$$^\dagger$ & 200 & 2.54$^\dagger$ & 362 & 141 \\
    \hline
    Median & A & 40 & 400 & $5\times10^3$ & 10$^\dagger$ & 2.30 & 33 & 15 \\
           & B & 3 & 170 & $10^5$$^\dagger$ & 100 & 2.30$^\dagger$ & 83 & 16 \\
    \hline
    \end{tabular}
    \\
    \raggedright
    Column descriptions: (1) wind boost factor, (2) magnetic field in the SAM, (3) wind age, (4) ambient gas density at 100 pc, (5) CR index, (6) derived forward shock radius, and (7) derived reverse shock radius.\\
    $^\dagger$ denotes fixed values.
    Model~A assumes an EC-dominated scenario with fixed $n_0 = 10\ \mathrm{cm}^{-3}$, while Model~B assumes a \textit{pp}-dominated scenario with fixed $t_\mathrm{w} = 10^5$ yr and $q_\mathrm{CR}$ to an gamma-ray index observed by \textit{Fermi}-LAT.
    \label{tab:model_parameters}
\end{table*}

\section{Wind Parameter Calibration with GeV detected Seyfert galaxies}\label{sec:sample}

\subsection{Fermi-Detected Seyfert Galaxies}\label{sec:Fermi Seyferts}
We select five Seyfert galaxies with GeV gamma-ray detections as our calibration sample: the Circinus galaxy, NGC~4945, NGC~4151, NGC~1068, and GRS~1734--292. 
Their observed properties are summarized in Table~\ref{tab:Fermi_Seyfert galaxies}. 
Three of them (Circinus, NGC~4945, and NGC~1068) come from the cross-match  \citep{2021ApJ...916...28T} between the \textit{Swift}-BAT 105-month catalog \citep{2018ApJS..235....4O} and the \textit{Fermi}-LAT 4FGL-DR2 catalog \citep{2020arXiv200511208B}\footnote{
We do not include IGR~J13109-5552 because this object has recently been reported as a blazar \citep{2024ApJ...977...56P}.
}. 
These sources are classified as ``Seyfert'' in the BAT catalog. 
We additionally include NGC~4151 and GRS~1734--292, which are detected by \textit{Fermi}-LAT \citep[4FGL-DR4,][]{2023arXiv230712546B} and firmly established as Seyfert galaxies \citep[][respectively]{1943ApJ....97...28S, 1998A&A...330...72M}.
We consider that the selected five Seyfert galaxies, which do not have strong jets, are the best currently available laboratories to constrain CR acceleration in disk-wind systems. 
As our sample is biased toward gamma-ray-bright Seyfert galaxies, we regard the estimated disk-wind contributions to the CGB and CNB as upper bounds (see Section~\ref{sec:CBR} for details).

Bolometric luminosities, which we use to determine the shock dynamics and CR powers, are taken from multiwavelength studies, or derived using the $L_\mathrm{X}$--$L_\mathrm{AGN}$ relation
\begin{align}\label{eq:bolometric_correlation}
\log\left(\frac{L_\mathrm{AGN}}{\mathrm{erg\,s^{-1}}}\right)
=& 0.0378 \left[ \log\left( \frac{L_\mathrm{X}}{\mathrm{erg\ s^{-1}}} \right) \right]^2\\
&- 2.03 \log\left( \frac{L_\mathrm{X}}{\mathrm{erg\ s^{-1}}} \right) + 61.6.\nonumber
\end{align}
\citep{2004MNRAS.351..169M, 2017ApJ...835...74I}, 
where $L_\mathrm{X}$ is a hard X-ray luminosity at 14--195~keV \citep{2018ApJS..235....4O}.
    
Among these five sources, UFOs have been detected in NGC~4151 and NGC~1068, with velocities of $v_\mathrm{w} = 0.105c$ \citep[][see also \citealt{2025ApJ...988L..54X}]{2010A&A...521A..57T} and $0.280c$ \citep{2019ApJ...871..156M}, respectively. We adopt these values for these two galaxies. For the remaining galaxies, no UFO signatures have been detected so far\footnote{Since UFO detectability depends on the line of sight \citep[e.g.,][]{2000ApJ...543..686P, 2016PASJ...68...16N}, non-detections do not necessarily imply the absence of winds.}. For these, we set $v_\mathrm{w} = 0.1c$.  

\subsection{Model Setup And Parameter Calibration} \label{sec:parameter}
    
We calibrate the wind model for each Seyfert galaxy using two scenarios, hereafter Model~A and Model~B, such that the model reproduces the observed radio and GeV gamma-ray fluxes. 
Our model contains several parameter degeneracies.
For example, the wind boost factor $b_\mathrm{w}$ and the ISM density at 100~pc $n_0$ strongly correlate with \textit{pp} luminosity (see Equation~\ref{eq:pp dependence}). 
Because only a few data points are available from \textit{Fermi}-LAT observations of a single system, while our model has many parameters, a Bayesian analysis is not able to break these degeneracies effectively (see Section~\ref{sec:CBR} for details). 
Therefore, only a subset of the parameters are treated as free.
Parameters specified in Tables~\ref{tab:Fermi_Seyfert galaxies} and \ref{tab:model_parameters} are fixed accordingly for each source, and all other parameters are fixed to the fiducial values in Table~\ref{tab:fiducial parameters}.
In both Models~A and B, we assume the same geometry and injection scheme; only the set of free parameters and the dominant process in the GeV band differ.
    
\begin{figure*}
    \centering
    \includegraphics[width=0.8\linewidth]{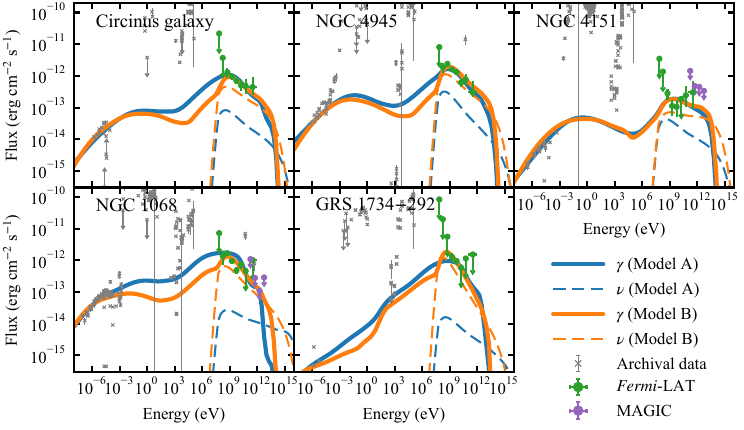}
    \caption{
    Multiwavelength SEDs of photons and neutrinos ($\nu+\bar{\nu}$ per flavor) of the Circinus galaxy, NGC~4945, NGC~4151, NGC~1068, and GRS~1734-292 from top-left to bottom right.
    In all panels, solid and dashed blue (orange) lines show total gamma-ray and neutrino fluxes in Model~A (B).
    Gray, green, and purple data points are archival data from \citet{https://doi.org/10.26132/ned1}, the 4FGL-DR4 catalog of \textit{Fermi}-LAT \citep{2023arXiv230712546B}, and 
    MAGIC \citep[][for NGC~4151 and NGC~1068, respectively]{2025arXiv250716527A, 2019ApJ...883..135A}.
    }
    \label{fig:SEDs FermiSy}
\end{figure*}

Each model reproduces radio and gamma-ray fluxes of all five Seyfert galaxies independently. 
In Model~A, we set the GeV gamma-ray emission to be dominated by EC scattering. We treat $b_\mathrm{w}$, $B_\mathrm{SAM}$, $t_\mathrm{w}$, and $q_\mathrm{CR}$ as free parameters, and fix the ambient density at $n_0 = 10~\mathrm{cm^{-3}}$, a typical value for ISM. 
On the other hand, in Model~B, we instead configure our model such that the GeV emission is dominated by \textit{pp} interactions. 
We then vary $b_\mathrm{w}$, $B_\mathrm{SAM}$, and $n_0$, while fixing the wind age to $t_\mathrm{w} = 10^5~\mathrm{yr}$, a typical AGN activity duration \citep[e.g.,][]{2015MNRAS.451.2517S}. 
The CR spectral index is set to the gamma-ray photon index measured by \textit{Fermi}-LAT \citep[4FGL-DR4,][]{2023arXiv230712546B}, since for \textit{pp} interactions the gamma-ray spectrum follows the parent proton slope closely \citep[e.g.,][]{2006PhRvD..74c4018K}.
The resulting parameter values and derived shock radii in both models are summarized in Table~\ref{tab:model_parameters}. Representative radio and gamma-ray SEDs for both models are shown in Figure~\ref{fig:SEDs FermiSy}. 

In Model~A, the GeV gamma-ray flux is mainly produced by EC scattering. The required $b_\mathrm{w}$ values span from 2 to 130, all below the maximum values inferred from recent XRISM observations \citep[$\sim10^3$,][]{2025ApJ...988L..54X}. The SAM magnetic field ranges from $B_\mathrm{SAM}\simeq59~\mu$G up to $\sim2~\mathrm{mG}$, and the wind ages range from $t_\mathrm{w}\sim2\times10^3$~yr to $10^5$~yr. The resulting radio, gamma-ray, and neutrino SEDs are shown in Figure~\ref{fig:SEDs FermiSy}. 
In some objects, \textit{pp} interactions provide a non-negligible contribution below GeV energies, but EC still dominates the GeV band.

In Model~B, the GeV gamma-ray flux is instead reproduced predominantly by \textit{pp} interactions. Here, $b_\mathrm{w}$ ranges from 1 to 130, $B_\mathrm{SAM}$ ranges from $17~\mu$G to $350~\mu$G, and the ambient density $n_0$ spans $75$--$200~\mathrm{cm}^{-3}$. 
We note that EC still contributes to the GeV gamma-ray flux in some cases but, by construction, the dominant GeV contribution in this scenario arises from \textit{pp} interactions.

\begin{figure*}[t]
    \centering
    \includegraphics[width=0.8\linewidth]{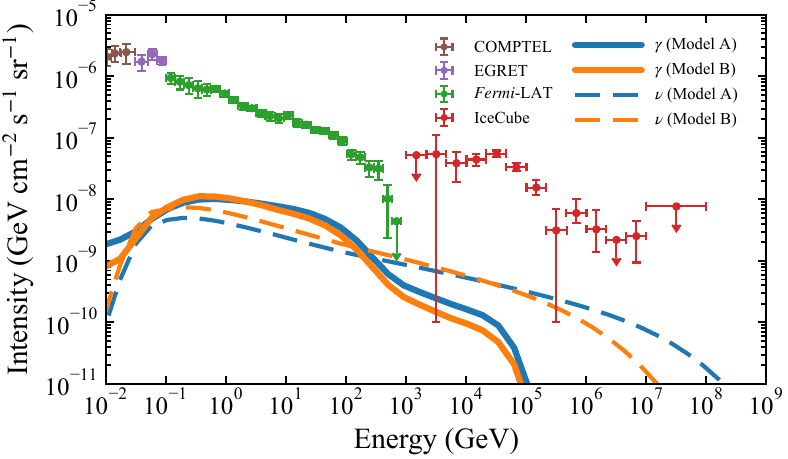}
    \caption{
    Cosmic gamma-ray (solid) and neutrino  (dashed) background intensities from AGN disk winds. 
    For neutrinos, we show the total emission, summed over all flavors. 
    Blue and orange curves correspond to our Model~A (leptonic dominated) and Model~B (hadronic dominated), respectively. 
    For the model parameters, we adopt the median values listed in Table~\ref{tab:model_parameters}, except for the wind age $t_\mathrm{w}$, where we invoke a uniform age distribution (see Equation~\ref{eq:CBR}). Brown, purple, green, and red points present the CGB spectrum measured by COMPTEL \citep{1996A&AS..120C.619K}, EGRET \citep{1998ApJ...494..523S}, \textit{Fermi}-LAT \citep{2015ApJ...799...86A}, and the CNB spectrum measured by IceCube \citep{2025arXiv250722233A, 2025arXiv250722234A}. 
    }
    \label{fig:CBR}
\end{figure*}

\section{Contributions of AGN Disk Winds to Cosmic Gamma-Ray and Neutrino Backgrounds}\label{sec:CBR}

We estimate the contribution of disk winds to the CGB and CNB by integrating the emission from individual sources over redshift, luminosity, and wind age. The resulting background intensity of gamma rays/neutrinos at an observed energy $E$ is 
\begin{align}\label{eq:CBR}
    I_{\gamma/\nu}(E)
    &= \frac{cf_\mathrm{w}}{4\pi}\frac{\Omega_\mathrm{w}}{4\pi~\mathrm{sr}} \int_{0.002}^5 \frac{dz}{(1+z)^2H(z)}
    \int_{43}^{48} d\log L_\mathrm{AGN} \nonumber \\
    &\quad \times \frac{d\Phi}{d\log L_\mathrm{AGN}}
    \frac{1}{t_\mathrm{Sal}-t_\mathrm{sf}}
    \int_{t_\mathrm{sf}}^{t_\mathrm{Sal}} dt_\mathrm{w} \nonumber \\
    &\quad \times 
    L'_{\gamma/\nu}(E', t_\mathrm{w}, L_\mathrm{AGN}) 
    \ \exp[-\tau_\mathrm{EBL}(z, E)] ,
\end{align}
where $f_\mathrm{w}$ is the fraction of AGNs hosting disk winds, $\Omega_\mathrm{w}$ is a representative opening solid angle for AGN disk winds, which acts as an anisotropic correction factor, $H(z)=H_0 \sqrt{\Omega_\mathrm{M}(1+z)^3 + \Omega_\Lambda}$ is the Hubble parameter with the Hubble constant $H_0$, $d\Phi/d\log L_\mathrm{AGN}$ is the AGN bolometric luminosity function, $t_\mathrm{Sal}$ is the Salpeter time, which is the time for an SMBH accreting with the Eddington luminosity to e-fold in mass \citep{2002MNRAS.335..965Y}, $t_\mathrm{sf}$ is the shock-formation time, and $L'_{\gamma/\nu}(E', t_\mathrm{w}, L_\mathrm{AGN})$ is the photon/neutrino luminosity of a single AGN disk wind at a rest frame energy $E'=(1+z)E$.

We fix parameters in Equation~(\ref{eq:CBR}) to reasonable values based on observations and simple estimations.
We set $f_\mathrm{w} = 0.4$, consistent with the observed fraction of AGNs hosting UFOs \citep{2010A&A...521A..57T}, and consider the limit $\Omega_\mathrm{w}=4\pi~\mathrm{sr}$. This limit ensures that our model yields an optimistic upper estimate of the disk wind contribution to the CGB and CNB, as the true value of $\Omega_\mathrm{w}$ would be less than this choice. 
For the AGN luminosity function $d\Phi/d\log L_\mathrm{AGN}$, we adopt the bolometric luminosity function of \citet{2014ApJ...786..104U} and multiply it by a factor of 1.5 to account for the Compton-thick population. 
We adopt the redshift range from 0.002 to 5 as in \citet{2014ApJ...786..104U}, below which local over-density effects are negligible. 
Following \citet{2016JCAP...12..012W, 2016NatPh..12.1116W, 2018ApJ...858....9L},  we set the Salpeter time to $t_\mathrm{Sal}=4\times10^7$~yr with a radiative efficiency of 0.1 \citep{2002MNRAS.335..965Y}. We approximate the shock-formation time as $t_\mathrm{sf} = R_\mathrm{launch}/c_\mathrm{s}$, where $R_\mathrm{launch}=100r_\mathrm{g}$ is the wind-launched radius \citep[e.g.,][]{2015MNRAS.451.4169G} and $c_\mathrm{s} = \sqrt{5 k_\mathrm{B}T_\mathrm{disk} / (3\mu_\mathrm{H} m_p)}$ is the adiabatic sound speed near the disk.

For the wind model parameters other than $t_\mathrm{w}$, we adopt the median values obtained from the GeV gamma-ray calibrated Seyfert samples for both Model~A and Model~B, respectively (Table~\ref {tab:model_parameters}), unless otherwise noted. 
The median value of the wind velocities is $v_\mathrm{w}=0.1c$, which is determined by the assumed values for non-UFO-detected Seyfert galaxies (see Section~\ref{sec:Fermi Seyferts}).
For the wind age $t_\mathrm{w}$, we assume a uniform distribution between $t_\mathrm{sf}$ and $t_\mathrm{Sal}$ and evolve $R_\mathrm{FS}$ and $R_\mathrm{RS}$. With this weighting, older winds contribute more significantly to the background, because the hadronic luminosity increases with $t_\mathrm{w}$ (see Figure~\ref{fig:parameter dependence}).

Figure~\ref{fig:CBR} shows the contributions of AGN disk winds to the CGB and CNB contributions for Model~A and Model~B.
For both models, the disk winds provide comparable contributions to the CGB and CNB, with Model~A extending to slightly higher energies in both backgrounds. 
In the GeV gamma-ray band, Model~A and Model~B reach up to $\sim5\%$ and $\sim4\%$ of the observed CGB above 10~GeV, respectively. 
At $\gtrsim$~TeV energies, the gamma-ray intensity decreases rapidly owing to EBL absorption. 
For the CNB, our models yield fluxes well below the diffuse neutrino intensity measured by IceCube in the TeV band, while at $\sim$ 100~TeV energies it can reach a level of $\sim 6$\% and $\sim8\%$ of the observed CNB for Model~A and B, respectively. 

We treat our background estimates as upper bounds.
As mentioned in Section~\ref{sec:Fermi Seyferts}, our sample is biased toward gamma-ray-bright, nearby Seyfert galaxies, and consequently, the adopted parameter values in the background calculation (Equation~\ref{eq:CBR}) are optimistic.
Moreover, our chosen wind-opening solid angle of $\Omega_\mathrm{w}=4\pi~\mathrm{sr}$ would exceed any realistic wind geometry. 
We also note that low-luminosity Seyfert galaxies, which dominate the AGN number density, are included in our background calculation through the luminosity function. Because such faint AGNs are far more numerous than bright ones, their cumulative contribution is fully captured in our population synthesis rather than understated.

Since we use a uniform distribution for the wind age $t_\mathrm{w}$, the dominant contribution to the background arises from the older population (see Figure~\ref{fig:parameter dependence}). 
The adopted parameter pairs $(b_\mathrm{w}, n_0~[\mathrm{cm}^{-3}])$ are $(40, 10)$ and $(3, 100)$ for Model~A and B, respectively. 
Therefore, characteristic wind power and ambient density are similar across the two models. In both cases, hadronuclear emission contributes a substantial fraction of the disk-wind CGB, at the level of $\sim50\%$ (Model~A) and $\sim40\%$ (Model~B). 
As a result, Models~A and B yield similar CGB and CNB spectra. 
The larger $b_\mathrm{w}$ and smaller $n_0$ in Model~A lead to higher maximum particle energies (see the top-left and bottom-left panels of Figure~\ref{fig:parameter dependence}), which naturally explains the modestly higher-energy extension of the CGB and CNB.

Although we have calibrated our parameter choices with observed galaxies, degeneracies still exist between some of them. 
In particular, Equation~(\ref{eq:pp dependence}) leads to $L'_{pp,\gamma/\nu, \mathrm{SAM}}\propto b_\mathrm{w}^{3/4}n_0^{5/4}$ for $\beta=1$, indicating a strong degeneracy between $b_\mathrm{w}$ and $n_0$.
We confirmed that a Markov-Chain Monte Carlo fitting was not able to break this degeneracy due to insufficient data.
In this context, Model~A and Model~B can be interpreted as two representative limiting cases of the parameter degeneracy. 
Model~A corresponds to a case with a relatively large $b_\mathrm{w}$ and moderate $n_0$, whereas Model~B corresponds to a case with enhanced $n_0$ and a more moderate wind boost.
Importantly, even across these limiting cases, the predicted diffuse background contribution does not change significantly.
Therefore, our result that AGN disk winds are unlikely to be dominant contributors to the CGB and CNB remains robust against parameter degeneracies.

\begin{figure}[t]
    \includegraphics[width=\linewidth]{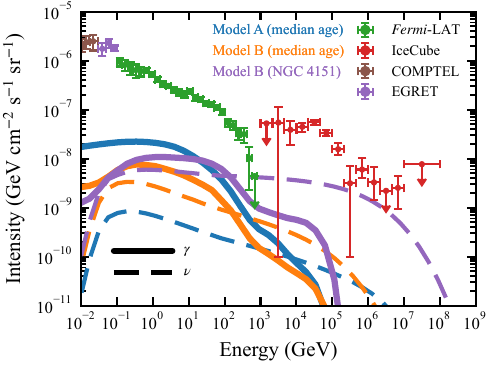}
    \caption{
    Same as Figure~\ref{fig:CBR}, but showing the cases with fixed $t_\mathrm{w}$ to the median values for Model~A (blue) and B (orange), and the case where parameters of NGC~4151 (purple) are adopted for all AGN disk winds with a uniform distribution of wind ages.}
    \label{fig:CBR_sub}
\end{figure}

To illustrate how the wind-age distribution and the adopted wind parameters affect our results, Figure~\ref{fig:CBR_sub} compares several representative cases. 
We show the CGB and CNB obtained by adopting the median parameter sets of Model~A and Model~B, respectively, while fixing the wind age to the median value of each model, and a Model~B realization where all disk winds share the parameters of NGC~4151 while retaining a uniform age distribution.

The NGC~4151-like case represents an extreme assumption where all AGN disk winds are assigned relatively hard CR spectra with $q_\mathrm{CR}=2.11$. 
In this situation, the disk winds could dominate the observed CNB in the relevant energy range. 
However, the other nearby GeV-detected Seyfert galaxies exhibit softer gamma-ray spectral indices (see Table~\ref{tab:model_parameters}), so an NGC~4151-like population cannot be representative of all AGN. 
A more realistic mixture of hard and soft sources, therefore, yields a disk-wind contribution to the CNB that remains subdominant (see Figure~\ref{fig:CBR}).

For Model~A, with the fixed age, the resulting neutrino background intensity is lower by a factor of $\sim 5$ compared to the case with a uniform age distribution, whereas for Model~B, fixing the wind age has only a modest impact. 
This difference mainly reflects the adopted ages, $t_\mathrm{w}=5\times10^3~\mathrm{yr}$ and $10^5~\mathrm{yr}$ for the median Model~A and Model~B cases, respectively. 
As discussed above, the background is dominated by older winds, so enforcing a relatively young age (as in the median Model~A case)  further suppresses the contributions to the CGB and CNB.

Although we have approximated the distribution of wind ages using uniform and $\delta$-functions in this work to illustrate the impact on our results, a more rigorous approach would be to model the age distribution of disk winds explicitly. 
At present, such modeling is not yet feasible. 
This is because the birth rates, death rates, and aging of disk wind systems are not yet well understood. 
However, these quantities may become accessible through future observations of disk wind systems using e.g., high spatial resolution radio telescopes like the Square Kilometre Array \citep{2009IEEEP..97.1482D}, which 
would support the development of such modeling in future extensions of this work.

\section{Discussion}\label{sec:discussion}
\subsection{GeV Emission Mechanisms in gamma-ray-detected Seyfert galaxies}\label{sec:gamma-ray origin}

In this work, we have modeled the GeV gamma-ray emission of five Seyfert galaxies in terms of shocks driven by AGN disk winds. Our lepto-hadronic wind framework can reproduce the observed radio and GeV fluxes with physically reasonable wind powers, magnetic fields, and ambient densities (see Figure~\ref{fig:SEDs FermiSy}). However, AGN disk winds are not the only plausible origin of the GeV emission from these systems. In principle, other components 
may make a major contribution --- for example, star-forming activity and/or weak jets. In the following, we discuss each object in our sample individually, comparing these alternative scenarios with the disk-wind interpretation.

\textbf{Circinus:} \citet{2013ApJ...779..131H} reported the detection of gamma-ray emission from the Circinus galaxy using four years of \textit{Fermi}-LAT data. 
The observed gamma-ray luminosity exceeds the level expected from its star-formation rate under a hadronic model and radio lobes under a leptonic model \citep[see also][]{2018MNRAS.474.4073W}. 
Using 10 years of \textit{Fermi}-LAT data, \citet{2019ApJ...885..117G} found that the gamma-ray luminosity could be broadly consistent with the calorimetric limit, but also reported marginal ($2.24\sigma$) evidence of year-scale variability, which may indicate the existence of AGN activity and is not expected for star-formation and radio-lobe emission scenarios. 

Although UFO features have not been reported from the Circinus galaxy, multi-wavelength observations have revealed multiphase outflows from sub-pc to kpc scales \citep{1998MNRAS.297.1202E, Greenhill2003ApJ...590..162G, Zschaechner2016ApJ...832..142Z, Izumi2018ApJ...867...48I}. 
The pc- to 10s-of-pc-scale outflow components are comparable to the forward and reverse shock radii predicted in our disk-wind model (see Table~\ref {tab:model_parameters}), which supports the interpretation that shocks in the AGN-driven wind could be a major contributor to the GeV emission in Circinus.

\textbf{NGC~4945:}
\citet{2010A&A...524A..72L} first reported the detection of GeV gamma-ray emission from NGC~4945 using 1.6 years of \textit{Fermi}-LAT data. The observed gamma-ray luminosity is close to the calorimetric limit of its starburst activity \citep{2010A&A...524A..72L,Fermi2012ApJ...755..164A, AguilarRuiz2021PhRvD.104h3013A}. 
However, several studies have argued for possible contamination by AGN activity \citep{Fermi2012ApJ...755..164A, 2016CRPhy..17..585O, 2018PASJ...70...49S}. 
In particular, \citet{2017ApJ...849...97W} reported a possible correlation between the X-ray and gamma-ray emission, suggesting the dominant gamma-ray emission region is located at a distance of $\sim10^3$--$10^4r_g$ from the SMBH. 

Although UFO signatures have not been reported from NGC~4945, multi-wavelength observations reveal multiphase outflows from pc to kpc scales, including Fe~K$\alpha$ emission extended over 10s to 100s of pc, a kpc-scale ionized outflow and hot wind, and disturbed dense molecular gas within $\lesssim100$~pc of the nucleus \citep{Marinucci2017MNRAS.470.4039M,Venturi2017FrASS...4...46V,Henkel2018A&A...615A.155H,PorrazBarrera2024ApJ...968...54P}. 
ALMA observations further resolve a dense rotating nuclear disk with a nearly coplanar molecular outflow within $\sim100$~pc, and gas densities of up to $n_0\sim10^{5}~\mathrm{cm^{-3}}$ have been inferred in its densest regions \citep{Henkel2018A&A...615A.155H}. 
These circum-nuclear structures on pc-to-100~pc scales are comparable to the shock radii in our disk-wind model (Table~\ref{tab:model_parameters}), supporting a scenario where shocks in the AGN-driven wind contribute substantially to the GeV emission. 
 
\textbf{NGC~4151:} 
Gamma-ray emission from NGC~4151 has been reported by some studies \citet[][although a separate analysis by  \citealt{Murase2024ApJ...961L..34M} reported a non-detection]{2023arXiv230712546B, 2025JCAP...07..013P}. Its modest star-formation rate is too low to drive the reported GeV luminosity alone. 
Given the detections of UFO features \citep[e.g.,][]{2010A&A...521A..57T,2015MNRAS.451.4169G}, \citet{2025JCAP...07..013P} proposed the gamma-ray emission could originate from a UFO-driven shock. 
They consider that shocks form where the AGN wind terminates in the surrounding medium, which can accelerate particles and account for a substantial fraction of the GeV flux. 
A collimated radio jet extending over $\sim300$~pc is known to exist \citep{2017MNRAS.472.3842W}, and jet emission from the nuclear region can also reproduce the observed radio and high-energy spectrum \citep{2023PASJ...75L..33I}. 
Although the current data do not allow us to disentangle the relative contributions of the disk wind and the compact jet, these results indicate that the GeV emission of NGC~4151 is predominantly associated with AGN-driven outflows.

Multi-messenger observations with IceCube have reported a $\sim3\sigma$ excess of TeV neutrinos from the direction of NGC~4151 \citep{Neronov2024PhRvL.132j1002N, IceCube2025ApJ...981..131A, IceCube2025ApJ...988..141A, IceCube2025arXiv251013403A}.  Although nearby blazars could potentially contaminate the signal, their contributions to the excess are likely minor \citep{Omeliukh2025A&A...694A.203O}. 
According to our work, an AGN-driven disk-wind would be difficult to account for the reported excess \citep[see also][]{2025JCAP...07..013P}. 
The most likely scenario at present is neutrino production in the AGN corona \citep[e.g.,][but see also \citealt{2023PASJ...75L..33I}]{Murase2024ApJ...961L..34M}.

\textbf{NGC~1068:}
Gamma-ray emission from the Compton-thick Seyfert galaxy NGC~1068 has been firmly established with \textit{Fermi}-LAT \citep{Fermi2012ApJ...755..164A}. Early works demonstrated that a pure starburst origin is disfavored, as the expected calorimetric output from the circumnuclear star-forming region could fall short of the observed GeV luminosity \citep{YoastHull2014ApJ...780..137Y,Eichmann2016ApJ...821...87E}. High-resolution ALMA observations reveal massive molecular outflows and a clumpy circumnuclear disk on scales of a few to several 100 pc \citep{Impellizzeri2019ApJ...884L..28I}, and a possible UFO feature has also been reported in X-rays \citep{2019ApJ...871..156M}. These findings motivate an AGN disk-wind scenario as the engine of the GeV emission \citep[][]{2016A&A...596A..68L}. A weak radio jet is known to exist in this system, but detailed modeling indicates that its contribution to the GeV flux is subdominant compared to the wind component \citep{Salvatore2024A&A...687A.139S}.

Neutrino observations with IceCube have further revealed a $\sim4.2\sigma$ excess at TeV energies from the direction of NGC~1068 \citep{IceCube2022Sci...378..538I}. 
A widely discussed explanation is hadronic emission from a compact AGN corona. This can account for the IceCube flux,  while strong internal $\gamma\gamma$ attenuation naturally suppresses the accompanying GeV--TeV gamma rays \citep[e.g.][]{Inoue2020ApJ...891L..33I,Murase2020PhRvL.125a1101M,Eichmann2022ApJ...939...43E}. In such coronal models, GeV gamma rays must be produced elsewhere --- for example, in the larger-scale AGN-driven wind or the circumnuclear starburst. Alternative scenarios aim to explain both the gamma rays and neutrinos simultaneously: a failed inner disk wind has been proposed as a common origin for the two messengers \citep{2022arXiv220702097I}, while hybrid models combining coronal and starburst components can also account for the multi-messenger data \citep{Eichmann2022ApJ...939...43E,Ajello2023ApJ...954L..49A}. 
Photodisintegration of nuclei in the jet could contribute to both gamma-ray and neutrino signals \citep[e.g.,][but see also \citealt{Das2024ApJ...972...44D}, who disfavor the photodisintegration scenario]{2025PhRvL.134o1005Y}.

Overall, current studies favor a picture where the GeV gamma-ray emission is mainly powered either by starburst activities or AGN-driven winds, while TeV neutrinos predominantly originate in the deeply embedded coronal/inner-wind region, although some degeneracy among these proposed interpretations remain.

\textbf{GRS~1734--292:}
GRS~1734$-$292 has been detected by \textit{Fermi}-LAT \citep{2023arXiv230712546B}. Multi-wavelength studies indicate that neither star formation nor jet activity is sufficient to account for the gamma-ray emission from this system \citep{2024ApJ...965...68M}. 
Although no firm detection of a UFO has been reported, X-ray spectroscopy reveals an Fe\,\textsc{xxv} absorption line from a warm outflow with $v\sim10^{3}~\mathrm{km~s^{-1}}$ \citep{Tortosa2017MNRAS.466.4193T}. Therefore, an AGN disk wind would provide a natural explanation for the GeV signal \citep{2025ApJ...980..131S}.

\subsection{Magnetic Fields in the Shocked Wind Region}\label{sec:B_SW}

In our model, the magnetic energy fraction in the SW region, $\epsilon_{B, \mathrm{SW}}$, is fixed at $10^{-5}$. This estimate is guided by physical considerations. 
Specifically, we considered a wind launched from a spherical surface around an accretion disk. In this 
simplified approach, the ratio of magnetic to kinetic energy density at the wind launching site can be parameterized as 
\begin{align}
\frac{B_\mathrm{disk}^2}{8\pi}
= \epsilon_{B, \mathrm{disk}}
\frac{\dot{K}_\mathrm{w}}{4\pi R_\mathrm{launch}^2 v_\mathrm{w}} \ ,
\end{align}
where $B_\mathrm{disk}$ is the magnetic field strength at the disk surface. 
Under conservation of magnetic flux, 
\begin{align}
    4\pi R_\mathrm{launch}^2B_\mathrm{disk} = 4\pi R^2B_{\mathrm{w}}(R), 
\end{align}
where $B_{\mathrm{w}}(R)$ is the wind magnetic field strength at a radius $R$. The ratio of magnetic to kinetic energy density is then estimated as
\begin{align}
\epsilon_{B, \mathrm{w}}(R)
&= \frac{B_{\mathrm{w}}^2(R)/8\pi}
{\dot{K}_\mathrm{w}/(4\pi R^2 v_\mathrm{w})} \nonumber \\ 
&= \left(\frac{R}{R_\mathrm{launch}}\right)^{-2}
\epsilon_{B, \mathrm{disk}} \ . \label{eq:epsilon_B,w}
\end{align}
Magnetohydrodynamic simulations coupled with CR transport demonstrated that the amplification of the ratio at strong shocks is maximally $\sim6\times10^2$ (see \citealt{2025arXiv251013946N}).
Therefore, we estimate $\epsilon_{B, \mathrm{SW}}$ as  
\begin{align}\label{eq:epsilon_B,SW}
    \epsilon_{B, \mathrm{SW}}\lesssim6\times10^{2}\left(\frac{R_\mathrm{RS}}{R_\mathrm{launch}}\right)^{-2}\epsilon_{B, \mathrm{disk}}.
\end{align} 
In the case of our fiducial parameter set ($R_\mathrm{RS}\sim100~\mathrm{pc}$, $R_\mathrm{launch}=100r_\mathrm{g}$, and $M_\mathrm{BH}=10^9~M_\odot$\footnote{10\% of the Eddington luminosity of this SMBH mass reproduces our fiducial bolometric AGN luminosity of $10^{46}$~erg~s$^{-1}$.}; see Table~\ref{tab:fiducial parameters}), the inferred value from Equation~(\ref{eq:epsilon_B,SW}) is $\epsilon_{B, \mathrm{SW}}\lesssim2\times10^{-6}\epsilon_{B, \mathrm{disk}}$. 
A value of $\epsilon_{B, \mathrm{SW}}=10^{-5}$ is therefore possible with $\epsilon_{B, \mathrm{disk}}\gtrsim5$, which would be feasible with magnetically-driven winds \citep[see e.g.,][]{2016A&A...589A.119C}.

Some studies have adopted significantly larger values of $\epsilon_{B,\mathrm{SW}}$, such as $\epsilon_{B,\mathrm{SW}}=0.05$ \citep{2023MNRAS.526..181P}. 
Because $\epsilon_{B,\mathrm{SW}}$ is highly sensitive to the reverse-shock radius through magnetic-flux conservation, a smaller $R_\mathrm{RS}$ naturally leads to a larger post-shock magnetic fraction. 
In particular, assuming compact, pc-scale reverse shocks with $\epsilon_{B, \mathrm{disk}}\gtrsim3$ could allow $\epsilon_{B,\mathrm{SW}}\sim0.05$ (see Equation~\ref{eq:epsilon_B,SW}), which, in turn, enables proton acceleration up to EeV energies. 
However, such a small radius of $\sim1~\mathrm{pc}$ would correspond to very young wind with $t_\mathrm{w}$ of a few hundred years in our dynamical model.

\subsection{Influence of ISM Density Profile on Disk-Wind Contributions}\label{sec:beta}

\begin{figure}
    \centering
    \includegraphics[width=\linewidth]{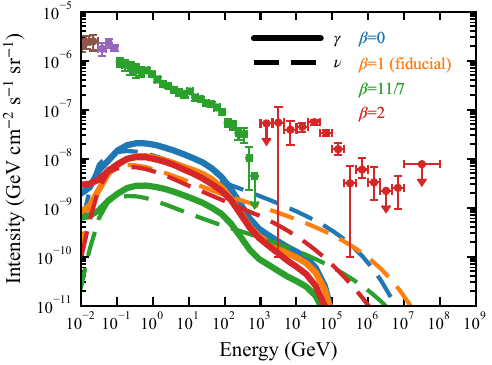}
    \caption{The dependence of the background intensities on the ISM index $\beta$. 
    We show the expected CGB and CNB intensities in the cases of $\beta=1$ (blue, our fiducial value), $\beta=0$ (orange), $\beta=11/7$ (green), and $\beta=2$ (red) for Model~B.
    The line styles and data points are the same as Figure~\ref{fig:CBR}.}
    \label{fig:CBR_beta}
\end{figure}

In this section, we investigate how the radial profile of ISM density $\beta$ affects the wind contributions to the background intensities. 
Figure~\ref{fig:CBR_beta} compares the background intensities for Model~B in the fiducial case ($\beta=1$) with the cases of $\beta=0$, 11/7, and 2. 
Our dynamical framework, which considers only the adiabatic phase (e.g., Equations~\ref{eq:ISM profile} and \ref{eq:FS radius}), cannot be applied for $\beta=2$ because an SW expands radiatively rather than adiabatically.
Therefore, we instead use Equation~(A3) in \citet[][$R_\mathrm{FS}\propto t_\mathrm{w}$]{1992ApJ...388..103K} to describe a forward-shock propagation for $\beta=2$ and calculate the background intensities.

For $\beta\leq11/7$, the predicted intensities decrease as $\beta$ increases, while for $\beta=2$, they become comparable to those for $\beta=1$.
The latter trend likely arises because in the $\beta=2$ case, the dominant contribution shifts toward the younger wind population, where shocks interact with denser gas and the \textit{pp} efficiency is consequently higher.
This non-monotonic behavior reflects different wind dynamics. 
For $\beta\leq11/7$, both the SAM and SW expand adiabatically, while the SW expands radiatively in the steeper case ($\beta>11/7$).

Despite the $\beta$ dependence, our result that disk winds are not dominant contributors to the background is robust to variations in the ISM profile choice.
Although $\beta=0$ provides the maximum contributions among the explored $\beta$ values as in Figure~\ref{fig:CBR_beta}, they are higher only by a factor of $\lesssim2$ than those of the fiducial case ($\beta=1$).
In addition, observations of the Milky Way halo \citep[e.g.,][]{2015ApJ...800...14M} suggest $\beta\approx1.5$, where the predicted intensities would be lower than the above cases.

\subsection{Comparisons with Previous Wind-based Background Models}\label{sec:comparison}

\begin{figure*}
    \centering
    \includegraphics[width=0.8\linewidth]{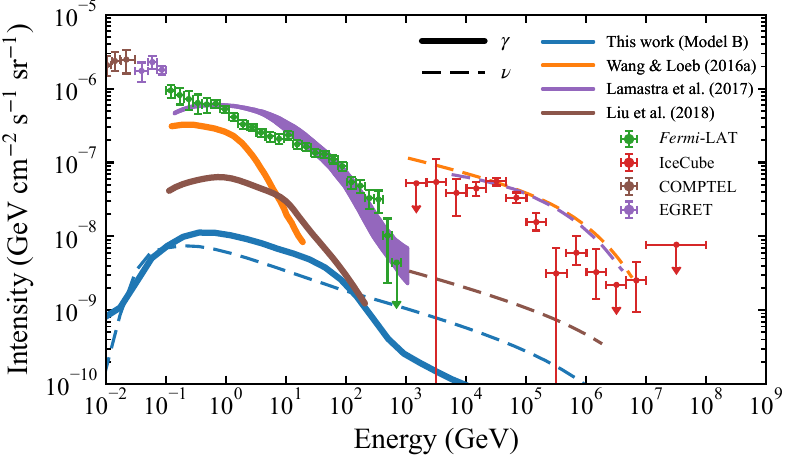}
    \caption{Comparison between the results obtained in this work and previous studies. 
    Thick solid and thin dashed lines show gamma-ray and neutrino intensities, respectively. 
    The blue lines correspond to Model~A in this study. 
    Other lines show the 
    background prediction by \citet[][orange]{2016JCAP...12..012W}, \citet[][purple]{2017A&A...607A..18L}, and \citet[][brown]{2018ApJ...858....9L}. 
    Differences in the intensities predicted by models can be attributed to different wind power, density, and luminosity function choices (see Section~\ref{sec:comparison} for details). 
    Green and red points show \textit{Fermi}-LAT data \citep{2015ApJ...799...86A} and IceCube data \citep{2025arXiv250722233A, 2025arXiv250722234A}.
    }
    \label{fig:CBR comparison}
\end{figure*}

Figure~\ref{fig:CBR comparison} compares our predicted CGB and CNB from AGN disk winds (blue curves; our Model~B) with previous wind-based calculations \citep{2016JCAP...12..012W, 2017A&A...607A..18L, 2018ApJ...858....9L}. 
As shown in Figure~\ref{fig:CBR}, Models~A and B produced similar CGB and CNB spectra. 
This similarity arises because hadronuclear emission forms a substantial fraction of the disk-wind CGB contribution in both models. 
Therefore, differences among studies are primarily controlled by the effective hadronuclear efficiency, which depends on the ambient gas density, CR confinement, and how the wind population is weighted over age.

To quantify this, it is useful to introduce the time-averaged \textit{pp} luminosity of an individual disk wind. 
Hereafter, $L'_{pp}$ denotes the \textit{pp} luminosity produced in the SAM, and we define
\begin{align}
    \bar{L}'_{pp}=\frac{1}{t_\mathrm{Sal}-t_\mathrm{sf}}\int_{t_\mathrm{sf}}^{t_\mathrm{Sal}}dt_\mathrm{w}L'_{pp,\gamma/\nu,\mathrm{SAM}}(t_\mathrm{w})\label{eq:time-averaged pp luminosity},
\end{align} 
which reduces to the thin-target scaling (e.g., Equation~\ref{eq:pp dependence}) and saturates in the calorimetric limit. 
For our fiducial Model~B background calculation, we obtain $\bar{L}'_{pp}\sim1\times10^{42}~\mathrm{erg~s^{-1}}$ for $L_\mathrm{AGN}=3\times10^{45}~\mathrm{erg\,s^{-1}}$. 
We also adopt a conservative UFO-hosting fraction of $f_\mathrm{w}=0.4$, whereas several previous works effectively assume ubiquitous winds ($f_\mathrm{w}=1$), which directly boosts the integrated background level by a factor of $\sim2.5$.

\citet{2016JCAP...12..012W} predict CGB and CNB intensities that are higher than ours by $\sim$one and two orders of magnitude, respectively. 
They solve wind dynamics hydrodynamically and adopt the AGN luminosity function of \citet{2007ApJ...654..731H}.
More importantly, they assume a denser and steeper circumnuclear gas distribution than we do, and they do not implement proton cooling. 
For instance, they model the distribution as the luminosity- and redshift-dependent density of $n_0\sim3\times10^{3}~\mathrm{cm^{-3}}$ at $100$~pc for $L_\mathrm{AGN}=3\times10^{45}~\mathrm{erg~s^{-1}}$, $z=0.1$, and $\beta=2^($\footnote{Although \citet{2016JCAP...12..012W} and \citet{2018ApJ...858....9L} consider broken power-law profiles for ISM densities, we adopt their assumed index at shock radii, where gamma-ray and neutrino luminosities peak.}$^)$. 
Under such a very dense environment compared to the case considered in Section~\ref{sec:beta}, unlike the behavior in Figure~\ref{fig:CBR_beta}, Equations~(\ref{eq:pp dependence}) and (\ref{eq:CBR}) show that \textit{pp} efficiency for $\beta=2$ is larger than that for $\beta=1$. 
Combining these assumptions strongly enhances the effective \textit{pp} efficiency and preferentially weights compact and young winds, for which the interaction rate is highest. 
Although they adopt a smaller wind-power normalization (e.g., $\dot{K}_\mathrm{w}/L_\mathrm{AGN}=0.03$, corresponding to $b_\mathrm{w}=0.6$ for $v_\mathrm{w}=0.1c$; Equation~\ref{eq:Kdot_wind}), substituting these values into Equation~(\ref{eq:time-averaged pp luminosity}) gives $\bar{L}'_{pp}\sim8\times10^{43}$~erg~s$^{-1}$ for $L_\mathrm{AGN}\approx3\times10^{45}$~erg~s$^{-1}$, which is $\sim80$ times larger than our fiducial value. 
Together with our conservative wind fraction of $f_\mathrm{w}=40\%$, this luminosity difference approximately accounts for the CNB discrepancy.
We also note that their neutrino-to-gamma-ray ratio appears larger than expected from simple hadronic calorimetry, suggesting possible tension in the implied energy budget as discussed by \citet{2018ApJ...858....9L}.
This tension may account for the CGB discrepancy.

Our predictions are also lower than those of \citet{2018ApJ...858....9L} by a factor of $\sim5$--10. 
As in \citet{2016JCAP...12..012W}, they solve the wind dynamics hydrodynamically and adopt 
the AGN luminosity function of \citet{2007ApJ...654..731H}. 
They set $\dot{K}_\mathrm{w}/L_\mathrm{AGN}=0.05$, corresponding to $b_\mathrm{w}=1$ for $v_\mathrm{w}=0.1c$ (Equation~\ref{eq:Kdot_wind}). 
They adopt similar ISM properties as in \citet{2016JCAP...12..012W}, but they normalize their luminosity-dependent density of $n_0=5\times10^2~\mathrm{cm^{-3}}$ at 100~pc for $L_\mathrm{AGN}=3\times10^{45}~\mathrm{erg~s^{-1}}$, which is 6 times less dense than \citet{2016JCAP...12..012W}. 
In this case, their \textit{pp} efficiency is close to unity (see their Figure~4) and Equation~(\ref{eq:time-averaged pp luminosity}) gives $\bar{L}'_{pp}\sim2\times10^{42}$~erg~s$^{-1}$. This is higher than our model by a factor of $\sim2$. 
The remaining difference is then explained by their larger effective wind incidence ($f_\mathrm{w}=1$ versus our choice of $f_\mathrm{w}=0.4$), together with differences in the adopted luminosity function and dynamical treatment. 
Overall, these effects can plausibly account for their factor of $\sim5$--10 higher background intensities.

By contrast, the results of \citet{2017A&A...607A..18L} remain difficult to reconcile quantitatively within our simplified comparison. 
Their model adopts a semi-analytic AGN evolution framework \citep{Menci2014A&A...569A..37M} and assumes a disk calorimetric treatment in which a geometry- and time-averaged fraction of CR protons interacts with dense gas. 
In addition, differences in the treatment of adiabatic losses and radiative transfer can alter the effective hadronic yield. 
These modeling choices plausibly contribute to their higher background intensities, but a detailed one-to-one comparison is beyond the scope of this paper.

\subsection{Future tests of the disk-wind scenario in UFO host galaxies}\label{sec:future}

To test the disk-wind scenario, observations of UFO-hosting Seyfert galaxies will be important with upcoming high-energy facilities. Instruments such as  
the Cherenkov Telescope Array Observatory \citep[CTAO;][]{2019scta.book.....C}, 
the Large Array of Imaging Atmospheric Cherenkov Telescopes at the Large High Altitude Air Shower Observatory site \citep[LHAASO-LACT;][]{2025ChPhC..49c5001Z}, the Southern Wide-Field Gamma-Ray Observatory \citep[SWGO;][]{2019BAAS...51g.109H}, 
IceCube-Gen2 \citep[][]{2021JPhG...48f0501A}, Tropical Deep-sea Neutrino Telescope \citep[TRIDENT;][]{2023NatAs...7.1497Y}, The Pacific Ocean Neutrino Experiment \citep[P-ONE,][]{2020NatAs...4..913A}, and The Trinity PeV Neutrino Observatory \citep{2019BAAS...51g..67O}
will offer a step change in sensitivity and resolving capability compared to current instruments. 
These advances will expand the sample of potential disk-wind host targets 
that are within observational reach, and provide the opportunity to firmly distinguish between different emission scenarios in individual systems. 

\begin{figure*}
    \centering
    \includegraphics[width=0.9\linewidth]{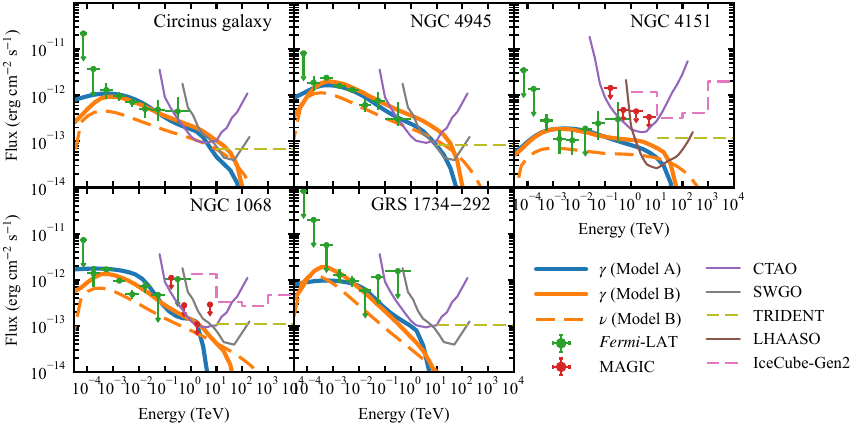}
    \caption{
    Same as Figure~\ref{fig:SEDs FermiSy} but focusing on the TeV energy band. 
    Over-plotted in each panel are the differential sensitivities of CTAO (purple), SWGO (gray), LHAASO-LACT (brown), IceCube-Gen2 (pink dashed), and TRIDENT (olive dashed), as labeled. 
    }
    \label{fig:TeVSEDs FermiSy}
\end{figure*}

For the five \textit{Fermi}-detected Seyfert galaxies analyzed in Section~\ref{sec:sample}, our calibrated models indicate that some are expected to be detectable with future observatories. 
Figure~\ref{fig:TeVSEDs FermiSy} presents the predicted SEDs in the TeV band for both a leptonic (Model A) and hadronic (Model B) emission scenario, together with sensitivity curves of CTAO, SWGO, TRIDENT, LHAASO, and IceCube-Gen2. This shows that the Circinus galaxy and NGC~4945 would both be detectable with CTAO with an exposure of $\sim 50$~hr, while NGC~1068 may also be detectable.
With $500$~hr of observations, LHAASO-LACT would be able to detect NGC~4151. 
In addition, SWGO could detect the Circinus galaxy and NGC~4945. 
In gamma-rays, the detection prospects between Model A and B are similar. 
However, neutrino emission could only arise under a hadronic scenario (Model~B). 
In the Circinus galaxy and NGC~4945, the predicted neutrino emission reaches the expected TRIDENT sensitivity in the multi-TeV range. 
TeV gamma-ray and neutrino detections of these nearby Seyfert galaxies, combined with existing GeV data, will be able to constrain the high-energy slope and cutoff of the gamma-ray component and provide direct constraints on $q_\mathrm{CR}$ and the relative EC and hadronic emission contributions from a disk wind.

\begin{table*}
    \centering
    \caption{Observed properties and model parameters of 11 UFO-hosting Seyfert galaxies.}
    \begin{tabular}{lcccc|lccccc}
    \hline \hline
    Object & $z$ & $L_\mathrm{X}$ & $L_\mathrm{AGN}$ & $v_\mathrm{w}$ 
           & Model & $b_\mathrm{w}$ & $B_\mathrm{SAM}$ & $t_\mathrm{w}$ & $n_0$ & $\log \epsilon_{B,\mathrm{SW}}$ \\
           & (1) & (2) & (3) & (4) 
           & (5) & (6) & (7) & (8) & (9) & (10) \\
           & -- & (erg\,s$^{-1}$) & (erg\,s$^{-1}$) & (c) 
           & -- & -- & ($\mu$G) & (yr) & (cm$^{-3}$) & -- \\
    \hline
    Ark~120 
      & 0.033  & $1.8 \times 10^{44}$ & $6.1 \times 10^{45}$ & 0.27 
      & A & 10 & 3.2 & $1.9 \times 10^{5}$ & 10$^\dagger$ & $-5$ \\
      &       &                          &                          &      
      & B & 3  & 20  & $10^5$$^\dagger$        & 250    & $-5$ \\
    Mrk~79 
      & 0.022  & $4.8 \times 10^{43}$ & $1.1 \times 10^{45}$ & 0.091 
      & A & 120 & 12.6 & $3.2 \times 10^{4}$ & 10$^\dagger$ & $-5$ \\
      &        &                         &                         &      
      & B & 30  & 20   & $10^5$$^\dagger$        & 150    & $-5$ \\
    Mrk~290 
      & 0.030  & $4.8 \times 10^{43}$ & $1.1 \times 10^{45}$ & 0.14 
      & A & 95 & 100 & $1.0 \times 10^{3}$ & 10$^\dagger$ & $-6$ \\
      &        &                         &                         &      
      & B & 40 & 5   & $10^5$$^\dagger$          & 200    & $-5$ \\
    Mrk~509 
      & 0.034  & $2.8 \times 10^{44}$ & $1.1 \times 10^{46}$ & 0.17 
      & A & 6 & 16 & $6.0 \times 10^{4}$ & 10$^\dagger$ & $-5$ \\
      &        &                        &                        &      
      & B & 5 & 20 & $10^5$$^\dagger$          & 200    & $-5$ \\
    Mrk~766 
      & 0.013  & $9.8 \times 10^{42}$ & $1.6 \times 10^{44}$ & 0.091 
      & A & 300 & 400 & $3.0 \times 10^{2}$ & 10$^\dagger$ & $-5$ \\
      &        &                         &                         &      
      & B & 70  & 20  & $10^5$$^\dagger$         & 150    & $-5$ \\
    Mrk~841 
      & 0.036  & $1.0 \times 10^{44}$ & $3.0 \times 10^{45}$ & 0.034 
      & A & 200 & 80 & $4.0 \times 10^{3}$ & 10$^\dagger$ & $-6$ \\
      &        &                        &                        &      
      & B & 80 & 10 & $10^5$$^\dagger$          & 200    & $-5$ \\
    NGC~3783 
      & 0.0097 & $3.6 \times 10^{43}$ & $7.9 \times 10^{44}$ & 0.013 
      & A & 88 & 100 & $6.0 \times 10^{3}$ & 10$^\dagger$ & $-5$ \\
      &        &                        &                        &      
      & B & 40 & 35 & $10^5$$^\dagger$          & 100    & $-5$ \\
    NGC~4507 
      & 0.012  & $5.8 \times 10^{43}$ & $1.4 \times 10^{45}$ & 0.18 
      & A & 8 & 70 & $3.0 \times 10^{4}$ & 10$^\dagger$ & $-5$ \\
      &        &                        &                        &      
      & B & 2 & 80 & $10^5$$^\dagger$          & 200    & $-5$ \\
    NGC~7582 
      & 0.0053 & $4.9 \times 10^{42}$ & $6.7 \times 10^{43}$ & 0.26 
      & A & 30 & 500 & $1.0 \times 10^{3}$ & 10$^\dagger$ & $-5$ \\
      &        &                         &                         &      
      & B & 10 & 55  & $10^5$$^\dagger$          & 100    & $-5$ \\
    ESO~323--G77 
      & 0.015  & $1.6 \times 10^{43}$ & $2.7 \times 10^{44}$ & 0.0050 
      & A & 1000 & 20 & $2 \times 10^{5}$ & 10$^\dagger$ & $-7$ \\
      &         &                        &                        &      
      & B & 300 & 50  & $10^5$$^\dagger$         & 300    & $-5$ \\
    ESO~434--40 
      & 0.0085 & $3.4 \times 10^{43}$ & $7.2 \times 10^{44}$ & 0.12 
      & A & 10 & 9 & $5.0 \times 10^{4}$ & 10$^\dagger$ & $-5$ \\
      &        &                       &                       &      
      & B & 2  & 25 & $10^5$$^\dagger$         & 200    & $-5$ \\
    \hline
    \end{tabular}
    \\
    \raggedright
    Column descriptions: 
    (1) redshift \citep{https://doi.org/10.26132/ned1}; 
    (2) X-ray luminosity integrated from 14 to 195~keV \citep{2018ApJS..235....4O}; 
    (3) AGN bolometric luminosity derived using Equation~(\ref{eq:bolometric_correlation}); 
    (4) UFO velocity \citep{2010A&A...521A..57T}; 
    (5) model type (A: EC-dominated; B: \textit{pp}-dominated); 
    (6) wind boost factor; 
    (7) magnetic field in the SAM; 
    (8) wind age; 
    (9) ambient gas density at 100~pc; 
    (10) magnetic-energy fraction in the SW region. 
    $^\dagger$Model~A assumes an EC-dominated gamma-ray spectrum with fixed $n_0 = 10\ \mathrm{cm}^{-3}$, while Model~B assumes a \textit{pp}-dominated spectrum with fixed $t_\mathrm{w} = 10^5$~yr.
    \label{tab:UFO_combined}
\end{table*}

\begin{figure*}[t]
    \centering
    \includegraphics[width=0.9\linewidth]{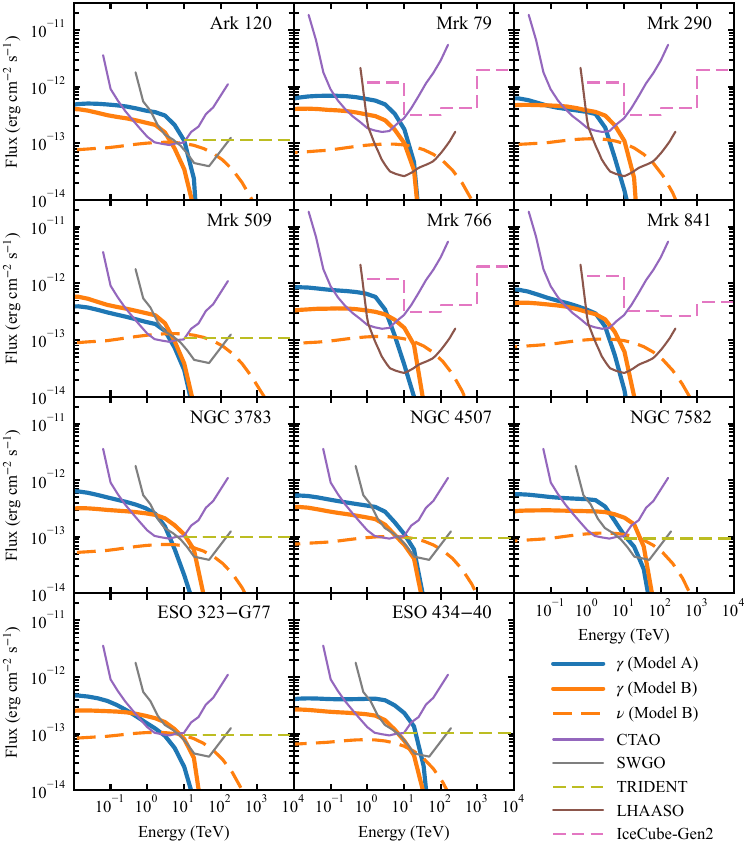}
    \caption{
    Same as Figure~\ref{fig:TeVSEDs FermiSy} but for UFO-hosting Seyfert galaxies.
    The adopted model parameters are summarized in Table~\ref{tab:UFO_combined}.}
    \label{fig:Prospect_SED}
\end{figure*}

We further explore the detection prospects for a broader sample of 11 UFO-hosting Seyfert galaxies, focusing on eleven objects with securely detected UFOs \citep{2010A&A...521A..57T}: Ark~120, Mrk~79, Mrk~290, Mrk~509, Mrk~766, Mrk~841, NGC~3783, NGC~4507, NGC~7582, ESO~323--G77, and ESO~434--40. Their basic observational properties are summarized in Table~\ref{tab:UFO_combined}. 
\citet{2024ApJ...968..116Y} has already shown that the radio spectra of these sources are consistent with synchrotron emission from primary electrons accelerated at forward shocks driven by disk winds, using essentially the same dynamical framework as adopted here. 

For each of these 11 Seyfert galaxies, we identify two representative parameter sets for which the disk-wind model reproduces the observed radio fluxes and yields TeV gamma-ray fluxes above the CTAO sensitivity, corresponding to EC-dominated (Model~A) and \textit{pp}-dominated (Model~B) cases. 
The treatment of model parameters follows the procedure described in Section~\ref{sec:parameter}, except for the magnetic-energy fraction in the SW, $\epsilon_{B,\mathrm{SW}}$, which we reduce for some objects to avoid overstating the observed radio fluxes. 
Specifically, under the Model A (EC-dominated) scenario, we adopt $\epsilon_{B,\mathrm{SW}}=10^{-7}$ for ESO~323--G77, and $10^{-6}$ for Mrk~290 and Mrk~841. 
These post-tuned values are consistent with weakly magnetized winds at the disk surface ($\epsilon_{B, \mathrm{disk}}\ll1$, see Section~\ref{sec:B_SW}).
These adjustments have little impact on the TeV gamma-ray and neutrino fluxes.
The resulting parameter sets are listed in Table~\ref{tab:UFO_combined}, and the corresponding TeV spectra are shown in Figure~\ref{fig:Prospect_SED}.

Figure~\ref{fig:Prospect_SED} indicates that, under our adopted parameter choices, all of the selected UFO-hosting Seyfert galaxies are promising TeV targets. In both Model~A and B scenarios, the predicted gamma-ray spectra for several objects lie above the CTAO sensitivity threshold over at least part of the 0.1--10~TeV range. 
At multi-TeV energies, the \textit{pp}-dominated spectra in Model~B using some sources remain within reach of wide-field gamma-ray observatories such as LHAASO and SWGO.
In the neutrino domain, the predicted fluxes for a subset of sources reach the expected sensitivity of TRIDENT, implying that future TRIDENT observations of UFO-hosting Seyfert galaxies could offer a direct and independent test of the hadronic disk-wind scenario.

X-ray spectroscopy will play a complementary and equally crucial role in constraining the wind parameters. 
Accurate measurements of $b_\mathrm{w}$ and $v_\mathrm{w}$ require high-resolution X-ray spectroscopy capable of resolving multiple absorption components. 
Recent XRISM results already suggest that disk winds in luminous AGNs often consist of multiple velocity components, with clumpy UFOs at characteristic velocities of $\sim 0.1c$ \citep[e.g.,][]{2025Natur.641.1132X}, and that typical values of $b_\mathrm{w}$ can be much larger than unity, $b_\mathrm{w}\sim 10$--$10^3$ \citep[e.g.,][]{2025Natur.641.1132X,2025ApJ...988L..54X}, substantially higher than values commonly adopted in earlier studies. 
Future systematic XRISM campaigns, together with next-generation facilities such as the Advanced Telescope for High-energy Astrophysics \citep[\textit{NewAthena},][]{2013arXiv1306.2307N},
will therefore be key to measuring the kinetic powers of disk winds in Seyfert galaxies. 
Combining X-ray constraints, GeV--TeV gamma-ray spectra (from CTAO and wide-field Cherenkov arrays), and neutrino limits/detections (from IceCube-Gen2 and TRIDENT) 
will allow stringent multi-messenger tests of whether AGN disk winds can accelerate CRs efficiently. This will firmly establish whether they 
can contribute appreciably to the high-energy gamma-ray and neutrino sky.

\section{Summary and Conclusions}\label{sec:conclusions}

In this work, we have revisited the role of AGN disk winds as sources of high-energy gamma rays and neutrinos in Seyfert galaxies, and as contributors to the cosmic gamma-ray background (CGB) and cosmic neutrino background (CNB). 
We developed a lepto-hadronic disk-wind framework where CRs are accelerated at the forward and reverse shocks driven by an AGN disk wind and we calibrated model parameters using nearby GeV-detected Seyfert galaxies.

We constructed a unified lepto-hadronic model for AGN disk winds where synchrotron, external-Compton, hadronic emission from \textit{pp}, and $p\gamma$ interactions (EC) emission in shocked ambient medium (SAM) shocked wind (SW) regions are treated self-consistently. 
Using this framework, we calibrated model parameters for five GeV-detected Seyfert galaxies by fitting their radio and GeV gamma-ray fluxes.
We considered two limiting scenarios: Model~A, in which the GeV emission is dominated by EC scattering in the SW region, and Model~B, in which the GeV emission is dominated by \textit{pp} interactions in the SAM.

For all five Seyfert galaxies, both models can reproduce the observed radio and GeV fluxes with physically plausible parameters. 
The inferred wind boost factors span $b_\mathrm{w}\sim\,$a few to $\sim 130$, in broad agreement with recent XRISM measurements of UFO kinetic powers, and the derived magnetic fields and ages are consistent with typical circum-nuclear environments. 
The corresponding forward- and reverse-shock radii are comparable to the scales of multiphase outflows and disturbed molecular gas seen in high-resolution observations, lending additional support to the disk-wind interpretation of the GeV emission in these objects.

We further used the Seyfert-calibrated parameters to evaluate the cumulative contribution of AGN disk winds to the CGB and CNB. 
We performed a population synthesis calculation, integrating the emission from disk winds over AGN bolometric luminosity and redshift, assuming a UFO-hosting fraction of $f_\mathrm{w}=0.4$, the bolometric luminosity function \citep{2014ApJ...786..104U}, and a uniform distribution of wind ages between the shock-formation time and the Salpeter time. 
With the median parameter sets from our GeV-calibrated Seyfert galaxies, we find that AGN disk winds can account for up to $\lesssim 5\%$ of the observed CGB above $\sim 10$~GeV, in both Model~A and Model~B scenarios. 
In the neutrino band, our models yield fluxes well below the diffuse neutrino intensity measured by IceCube at the TeV band, while around 0.1~PeV energies, it can reach the level of $\lesssim 10\%$ of the observed CNB in this framework. 
Considering that our adopted parameters are biased in favor of gamma-ray-bright Seyfert galaxies and we adopt a maximum possible opening angle (i.e., $\Omega_\mathrm{w}=4\pi~\mathrm{sr}$), we conclude that disk winds are unlikely to dominate either the CGB or the CNB. 

We further investigated the prospects for testing the disk-wind scenario with future gamma-ray and neutrino observatories. 
For the five \textit{Fermi}-detected Seyfert galaxies in our calibration sample, our models predict that the Circinus galaxy and NGC~4945 would be detectable with CTAO in both EC- and \textit{pp}-dominated scenarios for deep exposures, and that their hadronic neutrino emission in the \textit{pp}-dominated case reaches the expected TRIDENT sensitivity at multi-TeV energies. 
We also extended our analysis to eleven additional Seyfert galaxies with securely detected UFOs. 
For each of these objects, we identified EC-dominated and \textit{pp}-dominated parameter sets that reproduce the observed radio fluxes and yield TeV gamma-ray spectra above the sensitivity of CTAO. 
In several cases, the predicted multi-TeV gamma-ray emission remains within reach of wide-field facilities such as LHAASO and SWGO, while corresponding neutrino spectra approach the anticipated point-source sensitivity of IceCube-Gen2 or the stacking sensitivity of TRIDENT for a population of UFO-hosting Seyfert galaxies. 
These results indicate that TeV gamma-ray and neutrino observations of Seyfert galaxies will provide decisive tests of the disk-wind scenario and of the efficiency of CR acceleration in AGN cores. 

Our study is subject to several limitations that can be addressed in future work. 
In particular, we adopted simplified prescriptions for the wind age distribution, ambient density profile, and CR escape, and put focus on spherically averaged, one-zone representations of the SAM and SW regions. 
In reality, AGN winds are likely to be anisotropic and clumpy, and may interact with a multiphase circumgalactic medium in a way that cannot be fully captured by our self-similar description. 
In addition, we calibrated our model using a relatively small sample of nearby GeV-detected Seyfert galaxies, and the extrapolation of their median parameters to the entire AGN population carries uncertainties. 
Systematic XRISM campaigns and future X-ray missions such as \textit{NewAthena}, combined with deeper GeV--TeV gamma-ray and high-energy neutrino observations, will allow these assumptions to be tested and refined. 

Despite these caveats, our results support a picture where AGN disk winds can efficiently accelerate CRs and contribute non-negligible but subdominant fractions of the CGB intensity above $\sim 10$~GeV and the CNB intensity at 100~TeV for realistic choices of wind incidence, density, and energetics. 
Future joint analyses of Seyfert samples with X-ray, radio, GeV--TeV gamma-ray, and neutrino data will be crucial in establishing whether AGN disk winds are a ubiquitous, moderate contributor to the high-energy sky, or whether a subset of particularly efficient systems can play a larger role in shaping the diffuse extra-galactic gamma-ray and neutrino backgrounds.

\begin{acknowledgments}
The authors thank Susumu Inoue, Yutaka Fujita, Katsuaki Asano, and Kohei Ichikawa for useful comments and discussions. 
This research has made use of the NASA/IPAC Extragalactic Database (NED), which is funded by the National Aeronautics and Space Administration (NASA) and operated by the California Institute of Technology. 
N.S. was supported by JST SPRING, Grant Number JPMJSP2138, and is supported by JSPS Grant-in-Aid for JSPS Fellows Grant Number JP26KJ1621.
Y.I. is supported by JSPS KAKENHI grant No. JP22K18277 and JP26H00604.
ERO acknowledges support from the RIKEN Special Postdoctoral Researcher Program for junior scientists. 
He was also supported by a JSPS Postdoctoral Fellowship (KAKENHI Grant Number JP22F22327) while at The University of Osaka, where this work was initiated. 
\end{acknowledgments}

\software{astropy \citep{2013A&A...558A..33A,2018AJ....156..123A,2022ApJ...935..167A}
          }


    

\bibliography{sample701}{}
\bibliographystyle{aasjournalv7}



\end{document}